\documentclass[]{aa}  

\usepackage{graphicx}
\usepackage{subcaption}
\usepackage{txfonts}
\usepackage{xcolor}
\usepackage{float,subfloat}
\usepackage{CJKutf8}

\usepackage{orcid}

\definecolor{darkblue}{RGB}{0,60,134} 
\usepackage[colorlinks=true, linkcolor=darkblue, citecolor=darkblue, urlcolor=darkblue]{hyperref}

\newcommand{\micron}{\rm \mu m}
\newcommand{\kms}{\,\rm km\,s^{-1}}
\newcommand{\Msun}{\rm M_\odot}
\newcommand{\Ha}{{\rm H}$\alpha$\xspace}
\newcommand{\Nii}{[N\,{\sc ii}]\xspace}
\newcommand{\Hb}{{\rm H}$\beta$\xspace}
\newcommand{\Oiii}{[O\,{\sc iii}]\xspace}
\newcommand{\zspec}{z_{\rm spec}}
\newcommand{\zphot}{z_{\rm phot}}

\let\linenumbers\relax
\let\nolinenumbers\relax

\begin{document} 
    
   \title{RUBIES: A Spectroscopic Census of Little Red Dots}
   \subtitle{All Point Sources with V-Shaped Continua Have Broad Lines}

   \author{
        Raphael~E.~Hviding\orcid{0000-0002-4684-9005}\inst{\ref{i_mpia}}\thanks{hviding@mpia.de}
        \and Anna~de~Graaff\orcid{0000-0002-2380-9801}\inst{\ref{i_mpia}}
        \and Tim~B.~Miller\orcid{0000-0001-8367-6265}\inst{\ref{i_ciera}}
        \and David~J.~Setton\orcid{0000-0003-4075-7393}\inst{\ref{i_princeton}}\thanks{Brinson Prize Fellow}
        \and Jenny~E.~Greene\orcid{0000-0002-5612-3427}\inst{\ref{i_princeton}}
        \and Ivo~Labb\'e\orcid{0000-0002-2057-5376} \inst{\ref{i_swin}}
        \and Gabriel~Brammer\orcid{0000-0003-2680-005X}\inst{\ref{i_dawn},\ref{i_NBI}}
        \and Rachel~Bezanson\orcid{0000-0001-5063-8254}\inst{\ref{i_pitt}}
        \and Leindert~A.~Boogaard\orcid{0000-0002-3952-8588}\inst{\ref{i_leiden}}
        \and Nikko~J.~Cleri\orcid{0000-0001-7151-009X}\inst{\ref{i_psu}}
        \and Joel~Leja\orcid{0000-0001-6755-1315}\inst{\ref{i_psu}}
        \and Michael~V.~Maseda\orcid{0000-0003-0695-4414}\inst{\ref{i_wmad}}
        \and Ian~McConachie\orcid{0000-0002-2446-8770}\inst{\ref{i_wmad}}
        \and Jorryt~Matthee\orcid{0000-0003-2871-127X}\inst{\ref{i_ista}}
        \and Rohan~P.~Naidu\orcid{0000-0003-3729-1684}\inst{\ref{i_mit}}\thanks{NHFP NASA Hubble Fellow}
        \and Pascal~A.~Oesch\orcid{0000-0001-5851-6649}\inst{\ref{i_ge},\ref{i_dawn},\ref{i_NBI}}
        \and Bingjie~Wang~(\begin{CJK}{UTF8}{gbsn}王冰洁\end{CJK})\orcid{0000-0001-9269-5046}\inst{\ref{i_psu}}
        \and Katherine~E.~Whitaker\orcid{0000-0001-7160-3632}\inst{\ref{i_umass},\ref{i_dawn}}
        \and Christina~Williams\orcid{0000-0003-2919-7495}\inst{\ref{NOIRLab}}
    }
    \authorrunning{Raphael E. Hviding et al.\ }

   \institute{
    Max-Planck-Institut f\"ur Astronomie, K\"onigstuhl 17, D-69117 Heidelberg, Germany\label{i_mpia} \and 
    Center for Interdisciplinary Exploration and Research in Astrophysics (CIERA), Northwestern University, 1800 Sherman Ave, Evanston, IL 60201, USA\label{i_ciera} \and 
    Department of Astrophysical Sciences, Princeton University, 4 Ivy Lane, Princeton, NJ 08544, USA\label{i_princeton} \and
    Centre for Astrophysics and Supercomputing, Swinburne University of Technology, Melbourne, VIC 3122, Australia\label{i_swin} \and
    Cosmic Dawn Center (DAWN), Copenhagen, Denmark\label{i_dawn} \and
    Niels Bohr Institute, University of Copenhagen, Jagtvej 128, Copenhagen, Denmark\label{i_NBI} \and
    Department of Physics and Astronomy and PITT PACC, University of Pittsburgh, Pittsburgh, PA 15260, USA\label{i_pitt} \and
    Leiden Observatory, Leiden University, PO Box 9513, NL-2300 RA Leiden, The Netherlands\label{i_leiden} \and 
    Department of Astronomy \& Astrophysics; Institute for Computational \& Data Sciences; Institute for Gravitation and the Cosmos; The Pennsylvania State University, University Park, PA 16802, USA\label{i_psu} \and
    Department of Astronomy, University of Wisconsin-Madison, Madison, WI 53706, USA\label{i_wmad} \and
    Institute of Science and Technology Austria (ISTA), Am Campus 1, 3400 Klosterneuburg, Austria\label{i_ista} \and 
    MIT Kavli Institute for Astrophysics and Space Research, 70 Vassar Street, Cambridge, MA 02139, USA\label{i_mit} \and
    Department of Astronomy, University of Geneva, Chemin Pegasi 51, 1290 Versoix, Switzerland\label{i_ge} \and
    Department of Astronomy, University of Massachusetts, Amherst, MA 01003, USA\label{i_umass} \and
    NSF National Optical-Infrared Astronomy Research Laboratory, 950 North Cherry Avenue, Tucson, AZ 85719, USA\label{NOIRLab}
    }
    
   \date{}
 
  \abstract{
    The physical nature of Little Red Dots (LRDs) -- a population of compact, red galaxies revealed by JWST -- remains unclear. 
    Photometric samples are constructed from varying selection criteria with limited spectroscopic follow-up available to test intrinsic spectral shapes and prevalence of broad emission lines. 
    We use the RUBIES survey, a large spectroscopic program with wide color-morphology coverage and homogeneous data quality, to systematically analyze the emission-line kinematics, spectral shapes, and morphologies of $\sim$1500 galaxies at $z > 3.1$. 
    We identify broad Balmer lines via a novel fitting approach that simultaneously models NIRSpec/PRISM and G395M spectra, yielding 80 broad-line sources with 28 (35\%) at $z > 6$. 
    A large subpopulation naturally emerges from the broad Balmer line sources, with 36 exhibiting `v-shaped' UV-to-optical continua and a dominant point source component in the rest-optical; we define these as spectroscopic LRDs, constituting the largest such sample to date.
    Strikingly, the spectroscopic LRD population is largely recovered when either a broad line or rest-optical point source is required in combination with a v-shaped continuum, suggesting an inherent link between these three defining characteristics.
    We compare the spectroscopic LRD sample to published photometric searches. 
    Although these selections have high accuracy, down to $\rm F444W<26.5$, only 50-62\% of the RUBIES LRDs were previously identified.
    The remainder were missed due to a mixture of faint rest-UV photometry, comparatively blue rest-optical colors, or highly uncertain photometric redshifts.
    Our findings highlight that well-selected spectroscopic campaigns are essential for robust LRD identification, while photometric criteria require refinement to capture the full population.
  }

   \maketitle
%

\section{Introduction} \label{sec:intro}

The James Webb Space Telescope \citep[JWST;][]{Gardner2023} has revealed a remarkable population of high-redshift sources with extremely red rest-optical colors. 
These sources span a broad redshift range \citep[$z\sim1-10$;][]{deGraaff2024d} and show diverse morphological properties, ranging from unresolved point sources to large disks that extend several kpc \citep[e.g.][]{Baggen2023,Furtak2023,PerezGonzalez2023,Gibson2024,Williams2024,Xiao2024}. 
Follow-up spectroscopy has revealed a variety in spectral properties. 
Although high equivalent width emission lines explain the red broad-band colors of some sources \citep[e.g.][]{Larson2023}, others indeed have red continua, which can be smoothly rising or show strong spectral breaks \citep[e.g.][]{Carnall2024,Cooper2024,Wang2024b}. 
Importantly, this diversity in properties also points to a mixture of physical interpretations, including (dust-reddened) star formation, evolved stellar populations, or emission from active galactic nuclei (AGN).

A peculiar subset of red sources are distinguished by their highly compact nature, and are commonly referred to as little red dots (LRDs).
While the term was originally coined by \citet{Matthee2024} to describe potential AGN with broad Balmer emission that appeared red and compact in JWST/NIRCam rest-optical imaging, its usage has since expanded. 
In particular, several independent searches aimed at identifying extremely massive galaxies within the first Gyr of the Universe relied on NIRCam photometry alone to select candidates  based on very red observed rest-optical colors ($\rm F277W-F444W\gtrsim1$), indicative of strong Balmer breaks, along with non-detections at wavelengths shorter than $1\,\micron$, consistent with a Lyman-$\alpha$ break at $z\gtrsim6$ \citep[][]{Labbe2023,Barro2024}. 
These searches uncovered a population of objects with so-called `v-shaped' continua: red in the rest-optical but blue in the rest-UV, many of which also exhibited point-like morphologies in the long-wavelength (LW) NIRCam bands and have since also been referred to as LRDs.

Follow-up spectroscopy with JWST/NIRSpec of individual sources has suggested an intriguing link between compact, red photometric sources and broad-line AGNs. 
The strongly-lensed source with a v-shaped broad-band spectral energy distribution (SED) of \citet{Furtak2023} was shown to have broad ($\rm FWHM\sim3000\,\kms$) Balmer emission lines consistent with AGN emission \citep[][]{Furtak2024}. 
Among the candidate massive galaxies at $\zphot\sim7-9$ identified by \citet{Labbe2023}, one was spectroscopically confirmed as a broad-line AGN at $\zspec=5.6$ \citep{Kocevski2023}, while others were verified at high redshift and showed both strong Balmer breaks and broad Balmer lines \citep[][]{Wang2024b}. 
\citet{Greene2024} conducted the first systematic spectroscopic follow-up of photometric LRDs from \citet{Labbe2023b} as part of the UNCOVER survey \citep{Bezanson2024}, finding that 9/12 (75\%) sources exhibit broad Balmer emission. 
Collectively, these studies suggest a high incidence of broad-line AGN among some LRD samples, though the exact fraction appears sensitive to the selection criteria and highlights the need for uniform, large-scale spectroscopic follow-up.

If interpreted as AGN-dominated sources, the number density of LRDs would exceed that of the faint AGN expected from extrapolation of the quasar UV luminosity function by an order of magnitude \citep{Matthee2024,Pizzati2024}. 
Furthermore, if the LRDs were AGN with properties consistent with their lower-redshift  counterparts, the implied high black hole masses appear to greatly exceed local BH-galaxy scaling relations \citep[e.g.,][]{Harikane2023,Maiolino2023,Furtak2024,Kokorev2023}, though this may be, in part, due to biases in estimating host properties, AGN attenuation or bolometric luminosities, and/or black-hole masses \citep[e.g.][]{Li2024,rusakovJWSTsLittleRed2025,chenDustBudgetCrisis2025}.
The population of LRDs may therefore have important implications for our understanding of the formation and growth history of supermassive black holes. 

On the other hand, if the rest-optical continuum is dominated by starlight, it would imply the presence of very massive galaxies in the first Gyr \citep[up to $10^{11}\,\Msun$ by $z\sim7-8$;][]{Labbe2023}, requiring extremely efficient star formation that pushes the boundaries of the maximum stellar mass growth possible in the $\Lambda$CDM model \citep{MBK2023}. 
Combined with their highly compact morphologies, it would also imply that LRDs are the densest stellar systems in the Universe \citep[][]{Baggen2024,Guia2024,Ma2024,deGraaff2025}, exceeding observations and theoretical expectations of the maximum densities in star clusters \citep[][]{Hopkins2010,Grudic2019}. 
However this is unlikely to be true for all LRDs, as some show Balmer breaks that are far stronger than can be produced with evolved stars alone \citep{deGraaff2025,Naidu2025}.

This uncertainty has motivated systematic searches aimed at quantifying the prevalence of LRDs and characterizing their population-wide properties, which so far have focused primarily on photometry \citep[e.g.,][]{Labbe2023b,Barro2024,Kokorev2024,Kocevski2024,Perez2024,Akins2024}.
Although the precise selection criteria used differ for each study, all require a red rest-optical continuum, most require that the rest-optical morphology is unresolved or very compact, and some additionally require that the continuum is v-shaped, i.e.\ a blue rest-UV continuum as well as a red rest-optical continuum. 
Notably, the inferred number densities and SED properties vary greatly depending on selection method and modeling choices. 
Whereas some favor an AGN-dominated interpretation of the SED, using the LRD population to quantify the AGN bolometric, luminosity and/or black hole mass functions \citep[e.g.][]{Labbe2023b,Kocevski2024,Kokorev2024}, others argue that the SEDs may be best-described by stellar populations and that LRDs therefore represent a class of dust-obscured, star-forming galaxies \citep[][]{Perez2024} or dust-obscured post-starburst galaxies \citep[e.g.][]{Labbe2023,Labbe2024,Williams2024,Wang2024b,wangRUBIESJWSTNIRSpec2025}. 
\citet{Hainline2025} also point out that a large fraction of sources may not have red continua, but that the broad-band colors may be boosted by strong emission lines. 
Finally, brown dwarfs can also appear to have v-shaped broad-band photometric SEDs and are necessarily point sources \citep[e.g.][]{Langeroodi2023, Greene2024, Hainline2024, Hainline2025}.

Although targeted spectroscopic follow-up of small LRD samples has revealed a high fraction of broad Balmer emission lines and v-shaped continua, it remains unclear how these findings extend to the broader photometric samples in the literature, whose selection criteria can vary significantly. 
For example, \citet{Perez2024} compiled spectra from various spectroscopic surveys for 18 sources in their photometric LRD sample and found that only 3 show broad Balmer emission, three times fewer than reported in the systematic follow-up by \citet{Greene2024}. 
These discrepancies have critical implications for the interpretation of the physical properties of LRDs.

To robustly link the spectral properties of LRDs to these large photometric samples therefore requires comprehensive follow-up spectroscopy of photometric candidates. 
The JWST/NIRCam grism has been demonstrated to be highly successful at determining robust redshifts and selecting broad Balmer emission lines \citep[e.g.][]{Matthee2024,Naidu2024,Sun2025,Lin2024,Lin2025}, but simultaneous coverage of forbidden lines such as \Oiii is still rare, making it difficult to rule out broadening from stellar feedback and outflows. 
Only JWST/NIRSpec can reveal the continuum shape and at the same time kinematically resolve multiple emission lines. 
However, such targeted follow-up often leads to a complex spectroscopic selection function, and robustly quantifying the fraction of photometrically-selected LRDs with broad Balmer lines and v-shaped continua is therefore challenging. 

The \emph{Red Unknowns: Bright Infrared Extragalactic Survey} (RUBIES; \citealt{deGraaff2024d}) was designed to deliver a large spectroscopic dataset with a well characterized selection function: RUBIES has observed a large number ($\sim$300) of red sources without morphological pre-selection, while at the same time sampling several thousand galaxies with a broad distribution in color space. 
In this paper, we use the full RUBIES dataset at $z > 3.1$ to robustly quantify the prevalence of (1) broad Balmer emission lines, (2) v-shaped continua, and (3) a dominant point source in the rest-optical.
As we will show below, a population of spectroscopic LRDs, i.e.\ those that meet all three criteria, naturally arises from the data, hinting that these features may be physically interlinked.

In Section~\ref{sec:data} we present an overview of the RUBIES survey and the relevant data used in this work. We describe our methodology for measuring typical LRD features in Section~\ref{sec:methods} while Section~\ref{sec:results} explores the relationship between these characteristics to construct a spectroscopic LRD definition. 
Section~\ref{sec:phot_sel} compares our results to existing photometric LRD selection techniques. 
Finally, we present our summary and discussion in Section~\ref{sec:summary}.
Where relevant, we assume a flat $\Lambda$CDM cosmological model with $\Omega_{\rm m}=0.3$ and $h=0.7$. All magnitudes are reported using the AB system \citep{Oke1983} and non-detections ($<1\sigma$) are reported as their $1\sigma$ upper limits.

\section{Data \& Spectroscopic Sample} \label{sec:data}

The JWST Cycle 2 program RUBIES (GO-4233; PIs: de Graaff \& Brammer) is a 60-hour spectroscopic survey with the NIRSpec microshutter array (MSA) that has observed $\sim$4500 high-redshift sources selected across $\sim$150\,arcmin$^2$ from JWST Cycle 1 NIRCam imaging programs. A detailed description of the observing strategy, parent catalogs and spectroscopic selection function, as well as the spectroscopic data reduction can be found in \citet{deGraaff2024d}. In this section, we provide a brief summary of key details relevant to this work.

\subsection{JWST Imaging} \label{sec:imaging}

The RUBIES targets were selected from two extragalactic deep fields, the Ultra Deep Survey (UDS) and Extended Growth Strip (EGS). Both fields have extensive photometric coverage from X-ray to radio wavelengths, and were central to the CANDELS and 3D-HST surveys \citep[][]{Grogin2011,Koekemoer2011,Brammer2012,Skelton2014}. Public JWST/NIRCam imaging was obtained for these fields as part of multiple programs in Cycles 1-3. 

For the EGS the majority of NIRCam imaging data comes from the Cosmic Evolution Early Release Science Survey (CEERS; GO-1345, PI: Finkelstein; \citealt{Finkelstein2025}). This imaging spans approximately $80\,$arcmin$^2$ \citep{Bagley2023} in 7 different NIRCam filters (F115W, F150W, F200W, F277W, F356W, F410M, F444W), with a $5\sigma$ point source depth of 28.6 mag at $4\,\micron$. In addition, the Cycle 1 program GO-2234 (PI: Ba\~nados; Khusanova et al. in prep.) obtained NIRCam/F090W imaging over the same footprint as CEERS. 

The UDS was one of two fields observed as part of the Public Release IMaging for Extragalactic Research Survey (PRIMER; GO-1837; PI: Dunlop). This survey used the same 8 filters as the programs in the EGS, reaching a comparable typical depth (see \citealt{Donnan2024}), but for a larger area of $\sim220\,$arcmin$^2$. Combined, these surveys therefore provide a homogeneous dataset for spectroscopic follow-up.

All imaging data were reduced with the \texttt{grizli} pipeline \citep{grizli}, the details of which are described in \citet{Valentino2023}. The resulting image mosaics have a pixel scale of $0.04\arcsec$ and are publicly available from the DAWN JWST Archive (DJA). The RUBIES targets were selected using version 7.2 of the DJA image mosaics; in this paper we will use version 7.2 for the UDS mosaics and version 7.4 for the EGS mosaics to perform morphological modeling. 

The RUBIES photometric parent catalog was constructed primarily from the public catalogs on the DJA, with photometry measured in circular apertures of diameter $0.5\arcsec$ and photometric redshift measurements from \texttt{eazy} \citep{Brammer2008}.
This catalog was thoroughly visually inspected to remove artifacts and stars and supplemented with a small number of candidate high-redshift ($z>6.5$) sources from the photometric catalogs of \citet{Weibel2024}. 

We caution that the photometry in this parent catalog does not account for the wavelength-dependent point spread function (PSF) of JWST. To measure consistent colors in this paper, we therefore cross-match our final spectroscopic sample, as described in Section~\ref{sec:spectra}, to the catalogs of \citet{Weibel2024}, who provide photometry measured from image mosaics that were matched in resolution to the F444W filter using empirical PSF models. We use circular aperture flux measurements with diameter of $0.5\arcsec$ to compute colors, and use the Kron flux measured in the F444W band as a measurement of the total flux. 

\begin{figure*}[ht!]
    \centering
    \includegraphics[width=\textwidth]{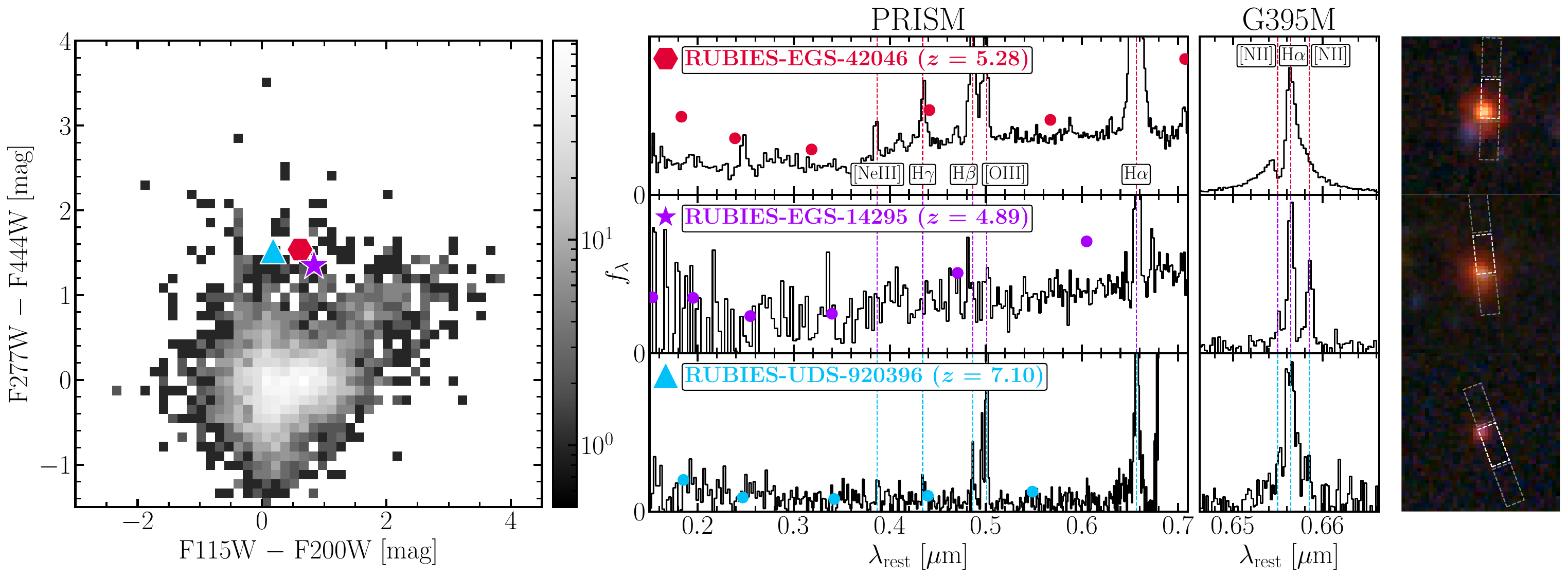}
    \caption{Diversity of red, high-redshift sources in RUBIES. Right: F115W$-$F200W vs.\ F277W$-$F444W for RUBIES with robust $\zspec>3.1$ (grey histogram) which populate a broad distribution in color space. Right: PRISM spectra and NIRCam photometry, G395M spectra (zoomed in on \Ha), and 1"$\times$1" NIRCam F444W/F277W/F150W RGB images for three RUBIES targets that are close in color space. The top row shows a typical LRD, with a v-shaped continuum, broad \Ha emission line and very compact morphology. The middle row shows an extended red object with a v-shaped continuum, but narrow \Ha and \Nii emission. The red source in the bottom row is a point source with a relatively blue continuum, but appears as red due to high EW emission lines, including a broad \Ha line. This demonstrates that sources with similar broad-band photometric colors can have very different spectral properties. }
    \label{fig:diversity}
\end{figure*}

\subsection{RUBIES Spectroscopy}  \label{sec:spectra}

The RUBIES dataset consists of 12 NIRSpec MSA pointings in the UDS and 6 pointings in the EGS, covering a total area of approximately $150\,$arcmin$^2$.
The pointing locations were chosen to optimize the total number of high-priority targets across the survey. 
Priorities were assigned to sources in the parent catalog using a small number of parameters (see \citealt{deGraaff2024d} for full details): the total flux at $4\,\micron$, $\rm F150W-F444W$ color, and photometric redshift.
The highest priority red sources were selected by requiring $\rm F444W<27$ and $\rm F150W-F444W>2$ with no restriction on photometric redshift; high-priority high-redshift sources were selected by $\rm F444W<27$ and $\zphot>6.5$ with visual vetting to remove low-redshift interlopers. 
The remainder of the catalog was rank-ordered using the source weight $W$, a quantity that is inversely proportional to the source number density in the parameter space of F444W, $\rm F150W-F444W$ and $\zphot$. 
In designing the masks, targets were assigned shutters in order of priority and weight. 

In total there are 4444 unique spectroscopic targets. 
As the result of the prioritization strategy, this sample includes a large number of rare, red sources at $z>1$ and the survey overall reaches high spectroscopic completeness ($>70\%$) in sparsely populated regions in observed color space. 
Source morphology was explicitly not used in any of the target prioritization, and RUBIES therefore probes a large variation of red sources, ranging from LRDs to dust-obscured extended star-forming disks. 
At the same time, the survey includes a representative sample of the high-redshift population that is less red, providing a crucial census sample to place rare sources in context. 

Of the 4444 targets, approximately 3000 were observed with both the low-resolution PRISM/CLEAR ($\mathcal{R}\sim50-500$; $0.6-5.5\,\micron$) and medium-resolution G395M/F290LP ($\mathcal{R}\sim1000-1500$; $2.7-5.5\,\micron$) disperser/filter combinations. The remainder were observed only with the G395M grating. Observations were taken by constructing 3-shutter slitlets for each target and performing a 3-point nodding pattern. The total exposure time is 48\,min for each disperser. 

The RUBIES spectra were reduced using version 3 of the \texttt{msaexp} pipeline \citep{msaexp}, as described in detail in \citet{Heintz2024} and \citet{deGraaff2024d}. These reductions offer two types of background subtraction for the PRISM mode. To ensure that the spectral extractions of the PRISM and G395M spectra are matched, we use only the local background subtracted spectra in this paper, i.e.\ those obtained from the image differences between the three nodded exposures. Spectroscopic redshifts were measured with the \texttt{msaexp} template fitting and visually inspected to assign quality flags.
Grading is fully described in Section 3.2 of \citet{deGraaff2024d}, however we note that a $\texttt{grade}=3$ corresponds to a robust redshift.
We note that RUBIES overall has a very high spectroscopic success rate, as approximately 90\% (70\%) of high-priority red sources (all sources) have robust redshifts.

For a subset of sources ($N=297$), we also use the newly-developed version 4 of \texttt{msaexp} (Brammer et al. in prep.).
The most important change in version 4 is in the flux calibration, which was re-derived empirically from standard stars observed in a range of commissioning and calibration programs. 
Crucially, these new calibrations extend the nominal wavelength ranges of the PRISM and G395M dispersers by $\pm0.2\,\micron$, enabling the measurement of \Oiii and \Ha emission down to $z\sim5.5$ and $z\sim3.1$ in G395M compared to $z\sim5.8$ and $z\sim3.4$ respectively, and \Ha out to $z\sim7.4$ in both dispersers compared to $z\sim7.1$.
We use these v4 reductions only for broad-line identification for sources at $\zspec \in[3.1,3.4]$ and $\zspec>6.9$, and rely on the v3 reductions for other analyses, i.e.\ continuum fitting. 

Lastly, we add one serendipitous compact red source ($\textrm{F444W}=24.1$) to our sample that was coincidentally observed as the neighbor of a lower-priority object (RUBIES-UDS-50432) in an outer slitlet.
Using a custom spectral extraction using the global background, we determine that the source is at $\zspec=6.42$ and could potentially be a bright LRD as preliminary visual inspections of its spectroscopy indicate a broad line and v-shaped continuum.
This source is referred to as RUBIES-UDS-57040 going forward.

To select our final sample for uniform and robust broad line identification, we require a robust  $\zspec$, i.e.\ visually inspected $\texttt{grade}=3$, and that either the \Ha or \Hb line fall into the G395M defined as any pixels falling within 1,000\,km\,s$^{-1}$ of the emission line position predicted by the DJA $\zspec$. 
This is especially relevant for robustly detecting broad lines at lower redshifts as the resolution of the PRISM disperser is very low ($\mathcal{R}<100$) at $<3\,\micron$.
In practice, this limits our sample to $\zspec>3.1$ with the extended coverage from the v4 reduction. 
This yields a sample of 1482 sources, with a median redshift of $\zspec=4.66$ and maximum redshift of $\zspec=9.3$, hereafter referred to as RUBIES with robust $\zspec>3.1$.
Of these 1198 (80\%) have PRISM spectroscopy.
As shown in Figures 5 and 6 of \citet{deGraaff2024d}, the $z\gtrsim3$ sources in RUBIES have a broad distribution in color space, spanning not only a broad range in the color used for target prioritization, $\rm F150W-F444W$, but also in $\rm F115W-F200W$ and $\rm F277W-F444W$. 

The diversity of our sample extends far beyond broad-band colors. In Figure~\ref{fig:diversity} we demonstrate that sources with similar observed colors can have drastically different spectral shapes, emission line properties, and emission line kinematics. 
In the top row, we present a prototypical LRD: a point-like red source with a v-shaped continuum and broad \Ha emission. 
The middle row shows a compact, yet resolved, dusty star-forming galaxy, which has a v-shaped continuum but narrow \Ha and \Nii emission. 
The bottom row shows an AGN, a point source with a relatively blue continuum that appears red in broadband photometry due to strong emission lines, including broad \Ha emission. 
Despite similar broad-band colors, the sources span a wide range of morphologies, star formation histories, dust content, and ionizing mechanisms. 
The spectroscopic diversity of these sources underscores the importance of the RUBIES spectroscopic dataset for disentangling the nature of red JWST sources.

\section{LRD Features} \label{sec:methods}

In this section we aim to robustly measure photometric and spectroscopic features commonly associated with LRDs: a broad Balmer line in Section~\ref{sec:broad}, a v-shaped continuum in Section~\ref{sec:cont}, and a dominant rest-optical point source in Section~\ref{sec:morph}.

\subsection{Broad Balmer Emission}\label{sec:broad}

Our goal is to robustly identify and measure broad emission components that are distinct from narrow line emission and uniquely associated with the Balmer lines, i.e.\ \Ha and \Hb.
This requires differentiating broad Balmer emission from potential contamination by other broadening mechanisms, such as galactic outflows that might similarly affect forbidden lines, e.g.\ \Oiii.

\begin{figure*}[ht!]
    \centering
    \includegraphics[width=\columnwidth]{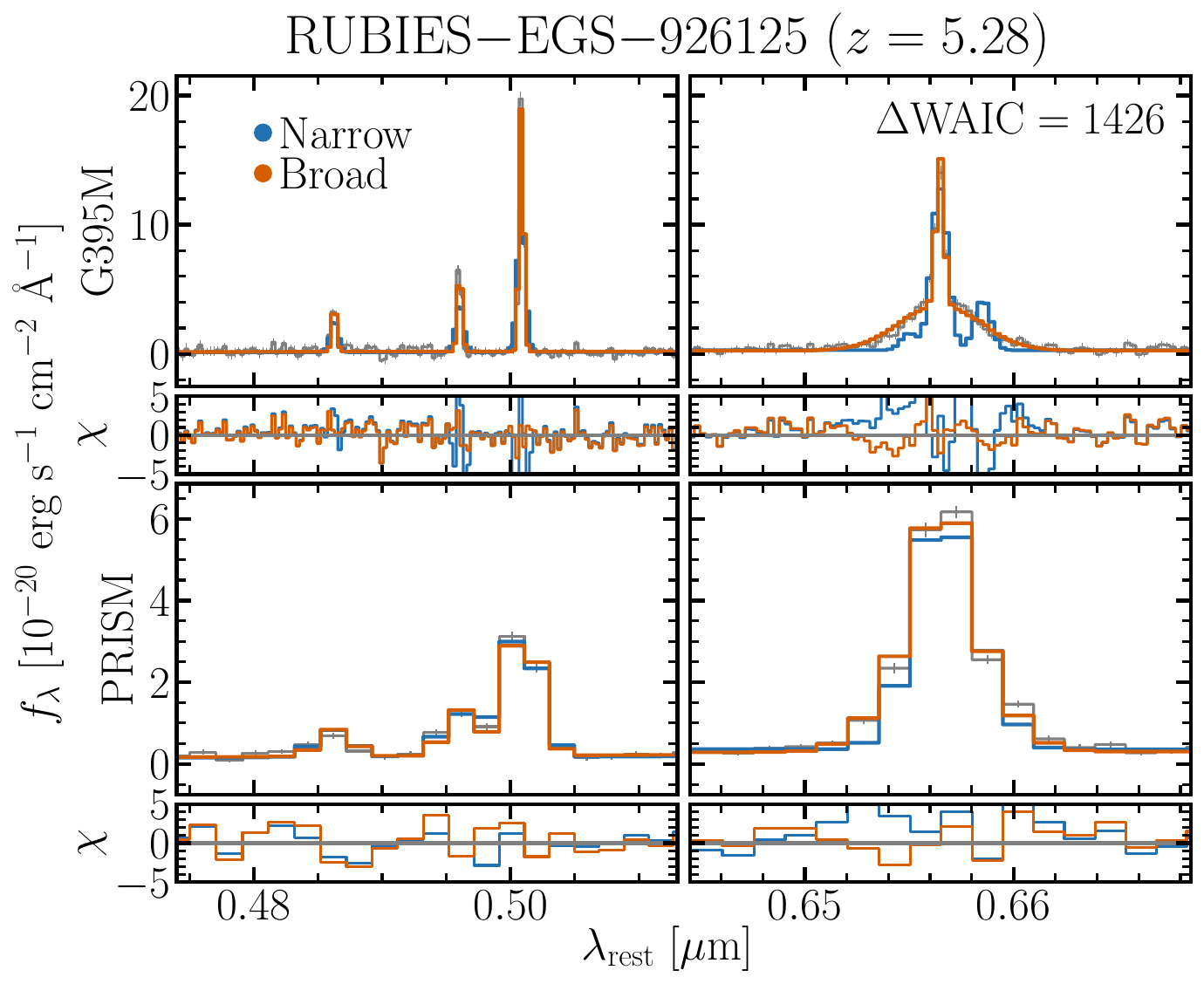}
    \includegraphics[width=\columnwidth]{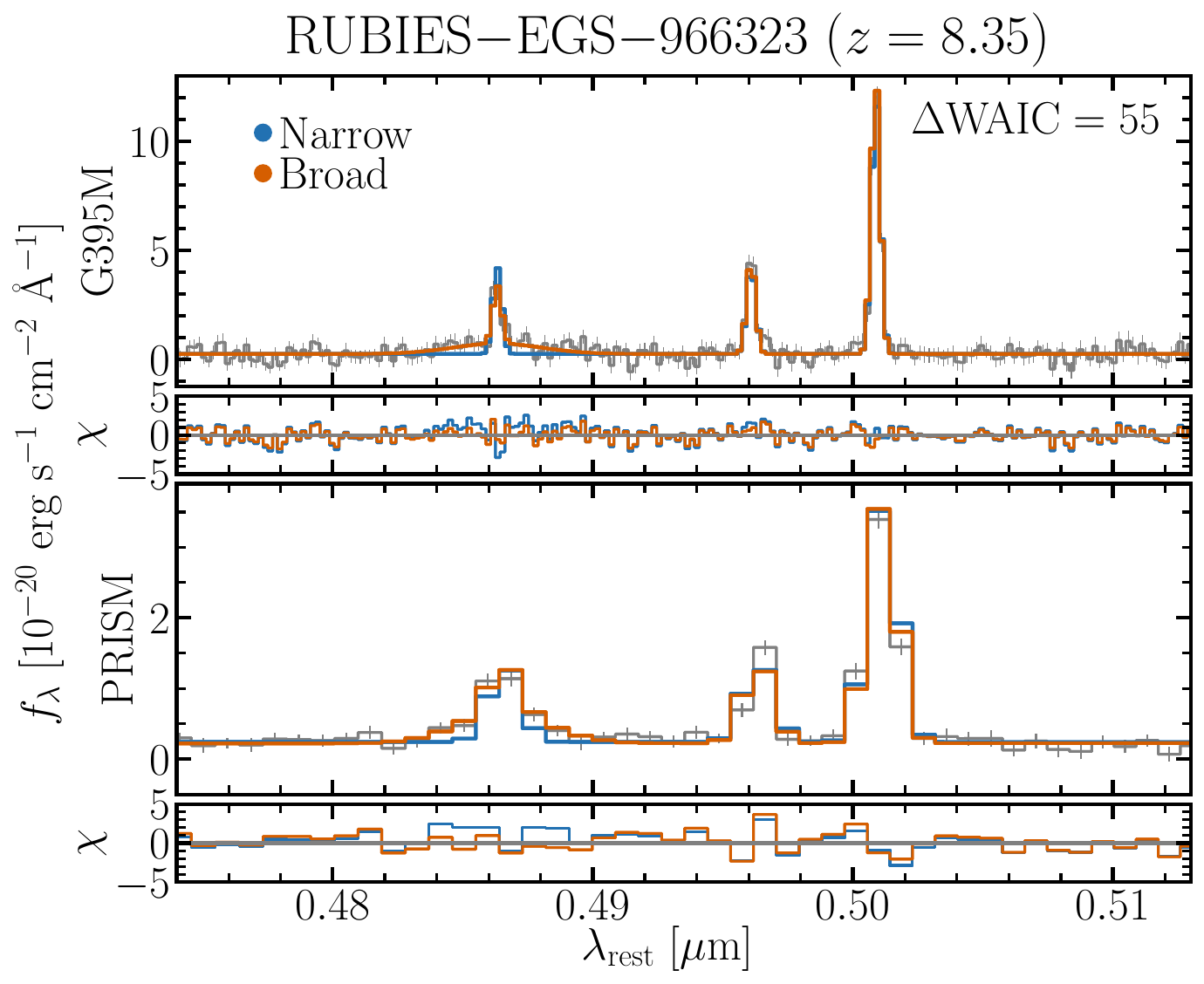}
    \caption{Zoom-ins of the spectroscopic fits for RUBIES-EGS-926125 (right) and RUBIES-EGS-966323 (left) using narrow (blue) and broad (orange) models for both G395M (top) and PRISM (bottom) spectra along with their residual deviations.
    We plot the maximum posterior sample from the MCMC fitting. 
    Simultaneous fitting leverages all available data to constrain linewidths in both gratings across different wavelength regimes.
    For 926125 we show both the \Hb$+$\Oiii and \Ha$+$\Nii complexes while for 966323 we only have coverage of the \Hb$+$\Oiii complex. 
    We note a broad component could not be conclusively fit in 966323 from the PRISM or G395M spectrum alone in \citet{Wang2024b} or \citet{Kocevski2024} but is detected in this work at the $>6.5\sigma$ level.}
    \label{fig:broad}
\end{figure*}

\subsubsection{Simultaneous Spectroscopic Fitting} \label{sec:simul}

To impose the strictest constraints on the presence of broad emission lines, we introduce the \texttt{Python} package \texttt{unite}\footnote{\texttt{unite} is hosted on \href{https://github.com/TheSkyentist/unite}{GitHub} with a v1 release coming soon.}, Uniform NIRSpec Inference (Turbo) Engine \citep{raphael_erik_hviding_2025_15585035}, to simultaneously fit all available NIRSpec spectra for each individual source within a given mask. 
Because extractions within a mask are performed using a consistent aperture, background subtraction, and trace extraction, we assume that the intrinsic source spectrum is the same across exposures. 
Differences between spectra arise primarily from the dispersion characteristics of the gratings, which affect the spectral resolution and sampling, as well as from relative calibration uncertainties between dispersers.
By jointly fitting these spectra, we leverage the complementary wavelength coverage, signal-to-noise (S/N), and resolution to enhance our sensitivity to broad emission features. 
For sources observed in multiple masks, we treat each mask independently, as differing apertures preclude the assumption of a shared underlying spectrum.

We begin by constructing a statistical model for emission lines and continua: we include the \Ha and \Hb emission lines along with the [O\,{\sc iii}]$\lambda\lambda4960,5008$\AA, [N\,{\sc ii}]$\lambda\lambda6549,6585$\AA, and [S\,{\sc ii}]$\lambda\lambda6718,6732$\AA\ doublets.
Our fitting region is centered around each line and extends $\pm$15,000\,km\,s$^{-1}$.
Overlapping regions are merged, resulting in two primary fitting windows: 4619\AA$-$5258\AA\ and 6222\AA$-$7070\AA\ in the rest frame. 
Each region is modeled with a linear continuum, where the slope $\theta$ follows a noninformative prior, $\theta \sim \mathcal{U}(-\pi/2, \pi/2)$, and the height is drawn from a broad uniform prior based on an initial continuum estimate.
Emission lines are modeled as Gaussians, with redshift $z \sim \mathcal{U}(\zspec - 0.005, \zspec + 0.005)$, FWHM $w_\textrm{narrow} \sim \mathcal{U}(0,750)$ km\,s$^{-1}$, and flux constrained to be positive, drawn from a broad uniform prior based on an initial flux estimate.

Because the default pipeline does not account for covariance in error propagation, we rescale the error spectrum in each fitting region to account for systematic over- or underestimation of uncertainties.
We mask $\pm$3,500\,km\,s$^{-1}$ around each expected emission line to isolate the continuum. 
The remaining region is fit with a weighted least squares (WLS) linear model from which we compute the reduced chi-squared, $\chi^2_\nu$. 
Assuming that the continuum is well described by a linear model, the normalized residuals should have unit variance. 
To enforce this, we multiply the error spectrum by $\sqrt{\chi^2_\nu}$ in each region and for each disperser. 
This correction ensures that uncertainties in the final fit reflect the true variance in the continuum, improving the reliability of emission line constraints.
Across all RUBIES spectra in this work, we find a typical correction factor of $\sim1.1\pm0.2$.

We build two physical models: a narrow model, consisting only of narrow emission lines, and a broad model, which includes additional broad Balmer emission lines.  
In the narrow model, all emission lines share a common redshift and intrinsic velocity width.  
The broad model extends this by incorporating two additional Gaussians for \Ha and \Hb.
To ensure reliable broad line detection, these additional lines are constrained to share the same redshift as the narrow lines.
However, their FWHM is drawn from the prior \( w_\textrm{broad} \sim \mathcal{U}(w_\textrm{narrow} + 100, 2500) \) km\,s$^{-1}$.  
For both models, the \Oiii and \Nii flux ratios are fixed to the quantum-mechanically derived values of 1:2.98 and 1:2.95 respectively \citep{Galavis1997}.

Fitting the physical model to each spectrum requires accounting for several instrumental and observational effects. 
First, emission lines are broadened by the line-spread function (LSF), using the NIRSpec LSF curves of an idealized point source obtained with \texttt{msafit} \citep{deGraaff2024a}. 
Because these are model LSFs rather than empirical calibrations, we introduce a scale factor drawn from a prior, \( s_\textrm{LSF} \sim \mathcal{N}(1.2, 0.1) \), truncated to [0.9, 1.5], to adjust for potential deviations.
To address systematic flux offsets between the G395M and PRISM spectra observed in \citet{deGraaff2024d}, we introduce a flux scale prior. 
The G395M scale is fixed at 1, while the PRISM scale follows \( s_\textrm{flux} \sim \mathcal{N}(1.1, 0.2) \), truncated to [0.5, 1.7], reflecting the observed range. 
Similarly, we account for pixel offsets between the two dispersers, with the G395M offset fixed at 0 and the PRISM offset drawn from \( \delta_\textrm{px} \sim \mathcal{U}(-0.3, 0.7) \) \textrm{px}.  
The choice of G395M as the reference is arbitrary and symmetric, ensuring that the relevant fluxes and redshifts remain recoverable for either disperser. 
Finally, to account for NIRSpec detector's undersampling of the LSF, the model is integrated in each pixel, rather than computed at the pixel center, before comparison with the observed data.

We implement these models using \texttt{NumPyro} \citep{phan2019}, a probabilistic programming library built on top of \texttt{JAX} \citep{jax2018github}. 
Leveraging \texttt{JAX}'s automatic differentiation and Just-In-Time (JIT) compilation, \texttt{NumPyro} allows us to efficiently define and sample from our joint Bayesian model. 
We employ Markov Chain Monte Carlo (MCMC) using the No-U-Turn Sampler \citep[NUTS;][]{hoffman2014} with 250 warmup steps followed by 500 posterior samples, ensuring robust convergence diagnostics and a typical runtime of $<$10 seconds for a typical source.
Finally, to compare the relative model fits, we compute the Widely Applicable Information Criterion (WAIC) \citep{Watanabe:WAIC}, which accounts for model complexity while estimating out-of-sample predictive accuracy. 

We present an example of our simultaneous broad and narrow fitting for RUBIES-EGS-926125 and RUBIES-EGS-966323 in Figure~\ref{fig:broad}.
The latter source was not conclusively identified as a broad-line source in \citet{Kocevski2024}, denoted as CEERS-9083, or \citet{Wang2024b}.
Due to the low S/N in the G395M spectrum and the relatively narrow width of the broad component, the broad-line was not identified from single spectrum fitting.
By simultaneously leveraging the resolution of the G395M to constrain the narrow line width and the S/N of the PRISM to detect deviations from the derived width, the broad line is measured with a $\rm \Delta WAIC = 55$, i.e.\ a $>6\sigma$ detection.

\subsubsection{Broad Line Validation} \label{sec:valid}

While a broad Balmer line will lead to a better fit with our fitting setup, data quality (DQ) issues can lead to a statistically improved fit even when the line profile is not well described by two kinematic components.
An initial examination of the data reveal several failure modes for our \texttt{unite} fits: spectral trace overlaps, low SNR, or large offsets between the PRISM and G395M dispersers.
To minimize contamination from false positives, we therefore implement the following quality cuts:
\begin{enumerate}
\item A $\Delta \text{WAIC}>11.8$, corresponding to a confidence level greater than $3\sigma$ is required, i.e.\ ensuring that the improvement in fit provided by the broad model is statistically significant over the narrow model alone.
    \item $w_\textrm{broad}>$1000\,km\,s$^{-1}$ to exclude narrow-line contamination or ambiguous features.
    \item In spectra with both \Ha and \Hb coverage, the flux of the broad \Ha must exceed the flux of the \Hb in order to eliminate unphysical fits.
\end{enumerate}

Following our quality cuts, three individuals (REH, AdG, JEG) visually inspected all 121 broad-line candidates. The inspectors identified 19 sources (15.7\%) with no clear evidence for a broad component (requiring 2/3 agreement), finding these cases showed flux contamination in the G395M due to other traces or DQ artifacts near the edge of the spectral range. 
We refer to these, and any other fitting failures due to DQ issues as indeterminate broad sources and present the common failure modes in Appendix~\ref{app:broad}.
In addition, we also denote objects without H$\alpha$ coverage and no broad Balmer detection as indeterminate broad sources.

However, a detected broad Balmer line does not guarantee that the line is not due to phenomena that can broaden other narrow lines, such as star-formation driven outflows. 
We therefore use information from forbidden transitions, such as \Oiii, to investigate the origin of the broad line.
For nearly every $\zspec \gtrsim 5$ galaxy we recover \Oiii in the G395M grating, provided the line does not fall into a chip gap or encounter other DQ issues.
To distinguish between these scenarios, we further refine our sample of 102 detected broad Balmer lines based on the following criteria:
\begin{enumerate}
    \item $w_\textrm{broad}\geq1500\kms$: we attribute the broadening to arise from non-stellar feedback origins and therefore be tied directly to the Balmer line itself ($N=69$). 
    While it is not impossible for feedback and outflows to generate velocities in excess of this limit, the number of these outflows even in AGN driven scenarios, drops sharply beyond $1000\kms$ \citep{haoActiveGalacticNuclei2005a,Leung2019,Forster2019}.  
    In addition, we note that we still visually inspect the forbidden-line properties where available for this sample and find no evidence for comparable broadening in these lines.
    \item $w_\textrm{broad}<1500\kms$: we then examine other strong nebular emission lines where available in the G395M grating, such as \Nii or \Oiii.
        \begin{enumerate}
            \item If no other narrow lines are detected, typically due to a lack of \Nii paired with redshifts where \Oiii falls out of the G395M spectrum ($z \lesssim 5$), the Balmer line is not classified as broad and is also referred to as an indeterminate broad source ($N=19$).
            \item If another narrow line is detected and there is equal broadening present in that line and/or significant residuals in its fit, as determined by 2/3 agreement, we do not consider the source to be a broad Balmer line source ($N=3$). 
            \item If another narrow line is detected and there is no broadening present, then the Balmer line is considered to be broad ($N=11$)
        \end{enumerate}
\end{enumerate}
We therefore conclude with a robust broad Balmer line sample of 80 galaxies and three with broadening in all narrow lines.
The availability of both sensitive PRISM and higher-resolution G395M grating spectra allows for robust broad-line selection, yielding one of the largest sample of broad-line objects to date at $z>4$ \citep{Harikane2023,Maiolino2023,Kocevski2023,Taylor2024,juodzbalisJADESComprehensiveCensus2025,Lin2024,Lin2025,zhuangNEXUSSpectroscopicCensus2025}.
The broad Balmer line sample is presented in Appendix~\ref{app:broad}, including examples of failures of the fitting due to DQ issues. 

Our emission line model, i.e.\ Gaussian profiles with uniform priors, is optimized for broad-line detection rather than detailed characterization. 
We caution that the reported FWHM values may not best represent the physical system, especially as LRDs often exhibit extended line wings that may be better modeled with Lorentzian functions \citep{wangRUBIESJWSTNIRSpec2025,deGraaff2025,Naidu2025,rusakovJWSTsLittleRed2025,Labbe2024}. 
Future work will address the full emission-line profile modeling taking these intricacies into account. 
\begin{figure}
    \centering
    \includegraphics[width=\columnwidth]{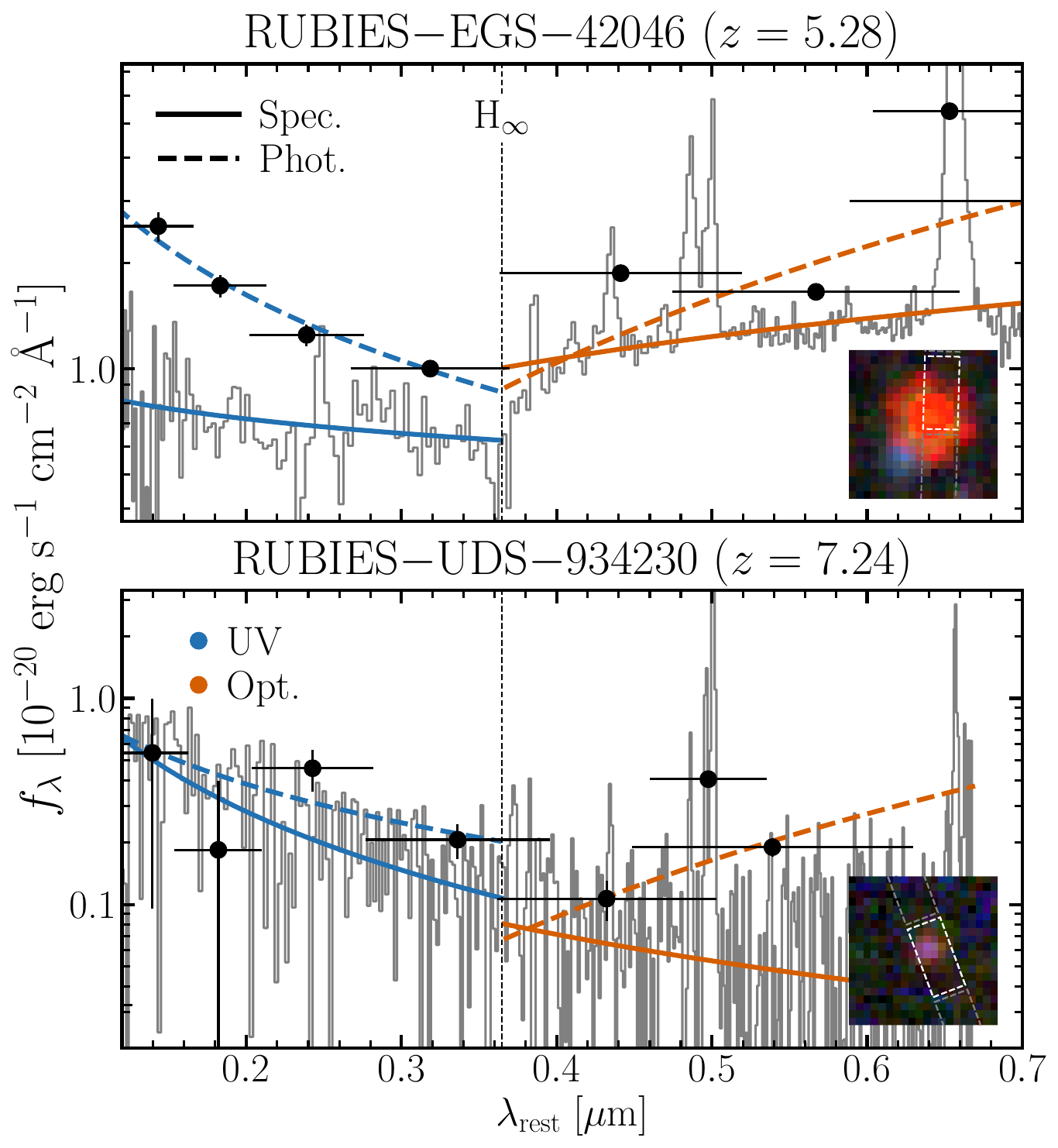}
    \caption{Rest-UV (blue) and rest-optical (orange) continuum fitting of RUBIES-EGS-42046 (top) and RUBIES-UDS-934230 (bottom) from PRISM spectroscopy (solid) and NIRCam photometry (dashed).
    We inset 0.5"$\times$0.5" NIRCam F444W/F277W/F150W RGB cutouts of the sources as well.
    We emphasize that spectroscopic rest-optical continuum fitting can mitigate the effect of strong emission lines but can suffer in S/N in the rest-UV where the spectrum is fainter or cannot capture extended rest-UV emission due to slit losses.}
    \label{fig:vshape}
\end{figure}

\subsection{V-Shaped Continuum} \label{sec:cont}

To robustly measure the continuum shape of LRDs, we adapt and extend the fitting method described in \citet{Setton2024} to measure spectral slopes from both PRISM spectroscopy and NIRCam photometry.
As established in \citet{Setton2024}, the break in LRDs preferentially occurs at the Balmer limit of 3645\AA\ (H$_\infty$) in a sample selected independent of broad-line width.
We therefore fix the continuum break location and fit the data on either side of the Balmer limit. 
For the rest-UV, we fit the range from 1200\AA$-$H$_\infty$, and for the rest-optical, we fit the range from H$_\infty-$7000\AA.
The continua ranges are fit using a power law of the form $f_\lambda = a \cdot \lambda_\textrm{rest}^{\,\beta}$ with non-linear least squares optimization. 

\subsubsection{Photometric Continuum}

We measure the continuum slopes from photometry by using all available wide NIRCam photometric bands whose central wavelength lie within the aforementioned spectral ranges. 
In addition, in order to fit a slope we require at least two photometric filters be available in the given spectral range.
As our goal is to use photometric continua to augment the spectroscopic measurements in the following section, we only measure a photometric continuum for sources in which we have PRISM spectroscopy available.

\subsubsection{Spectroscopic Continuum}

In order to mitigate the effect of strong emission lines, we mask the rest-frame spectrum $\pm50$\AA\ around the \Ha, \Hb, H$\gamma$, H$\delta$, He\,{\sc i}$\lambda\lambda4471,6680$\AA, [O\,{\sc ii}], and [Ne\,{\sc iii}] lines along with the \Oiii and \Nii doublets.
In addition, we require at least 25 wavelength elements be present after applying the emission line mask to fit a slope.
We use the photometric continuum fits to augment our detection of a blue rest-UV continuum due to the rest-UV faintness, however we always require a measurement of the rest-optical continuum from the spectroscopy. 
To define a source as v-shaped from spectroscopy, we impose the following criteria:
\begin{enumerate}
    \item A blue rest-UV continuum with a nonnegative fit: $\beta_\textrm{UV}\, < -0.2$ detected at the 2$\sigma$  level and $a_\textrm{UV}\, > 0$ from either spectroscopy or photometry.
    \item A red rest-optical continuum with a nonnegative fit: $\beta_\textrm{opt}\,\textrm{(Spec.)} > 0$ detected at the 2$\sigma$ level and $a_\textrm{opt}\,\textrm{(Spec.)} > 0$.
    \item $\beta_\textrm{opt} - \beta_\textrm{UV} > 0.5$ using the rest-UV slope from spectroscopy if it satisfies our blue rest-UV continuum cut in spectroscopy, otherwise from photometry.
\end{enumerate}
We are able to measure spectroscopic continua in 1158 (97\%) of our robust $\zspec>3.1$ sample with PRISM spectroscopy; of these 55 (5\%) are classified as v-shaped.

In Figure~\ref{fig:vshape}, we present the continuum fits for RUBIES-EGS-49140 and RUBIES-UDS-934230 using both photometric and spectroscopic data. 
Although both methods fit the data well, they differ in their results. 
While the photometry is often able to provide a higher S/N measurement, particularly in the rest-UV, it is significantly impacted by emission line contamination in broadband photometry, leading to different inferred red-optical slopes. 
However, photometry also can find bluer rest-UV slopes due to the extended UV `fluff' often seen in LRDs which can be lost in spectroscopy due to slit losses.

\subsection{Rest-Optical Point Source Morphology} \label{sec:morph}

Our morphological analysis addresses two key questions: (1) Is the source spatially resolved? and (2) if resolved, is there still a dominant point source component? 
We first assess basic resolvability in Section~\ref{sec:unr}, recognizing that Sérsic profile fitting inherently assumes extended emission and requires careful calibration against known point sources.
In addition, Section~\ref{sec:res} will investigate whether resolved sources still contain a dominant nuclear point source component.

We characterize unresolved sources by performing S\'ersic profile fits using \texttt{pysersic} \citep{Pasha2023} on all LW NIRCam bands, employing empirical PSFs from \citet{Weibel2024} for convolution. 
Our analysis requires S/N > 10 per filter and adopts uniform priors for the Sérsic index, $n\sim \mathcal{U}(0.65,6)$, and effective radius, $r_{\rm eff} \sim \mathcal{U}(0.25,25)$\,px.
The posterior distributions for the morphological parameters are sampled using the NUTS sampler implemented in \texttt{NumPyro} using 2 chains with 1,000 warm up and 1,000 sampling steps each.
We validate fits by requiring $\hat{r}<1.05$, effective sample size $<250$, and $\chi^2 /\ (\rm{\#\ px}) > 2$ to rule out poor sampling or large residuals.

\begin{figure}
    \centering
    \includegraphics[width=\columnwidth]{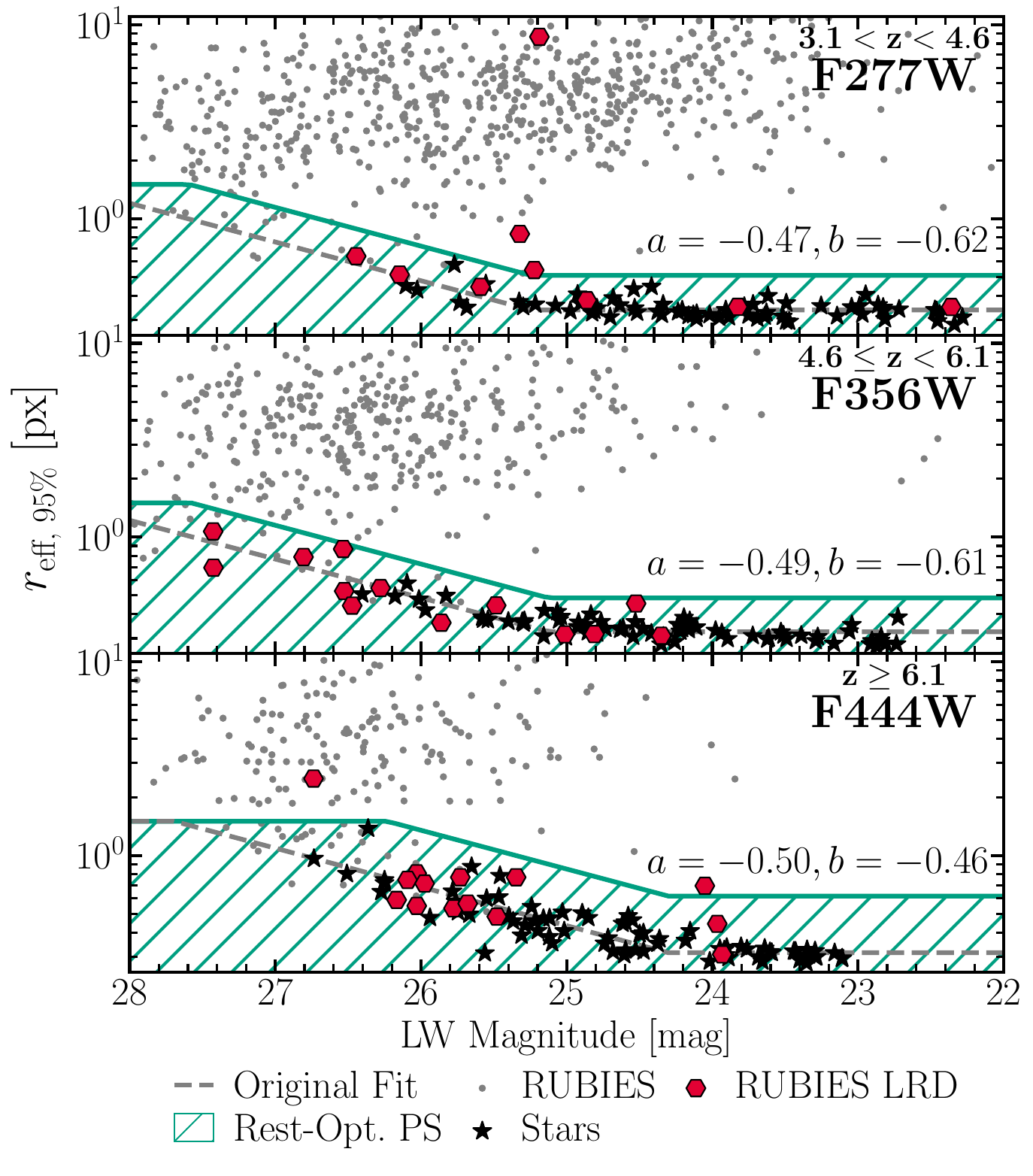}
    \caption{Stellar locus and morphological classification in LW NIRCam filters. Black stars show the 95th percentile effective radius posterior ($r_{\rm eff, 95\%}$) vs.\ magnitude for reference stars, with the best-fit stellar locus (grey dashed). We classify galaxies as point sources (green hashed) if they fall below the $+4\sigma_\textrm{resid}$ offset of this relation ($r_\textrm{rsv}$; green line). Grey circles show all RUBIES sources with robust $\zspec>3.1$ while red hexagons highlight spectroscopic LRDs (Section~\ref{sec:lrd}), plotted in the LW filter that traces the rest-5500\AA\ filter depending on their redshift.}
    \label{fig:morph}
\end{figure}

\subsubsection{Rest-Optical Unresolved Morphology}\label{sec:unr}

We first begin by assessing if the source is spatially resolved at all, i.e.\ can we robustly detect spatially extended emission beyond the PSF?
However, this is not straightforward from S\'ersic profile fitting, which by definition assumes an extended profile.
Even when a source’s inferred radius approaches the lower bound of the prior, this reflects an imposed modeling choice. 
While the prior limit can be adjusted, the interpretation ultimately depends on the precision of the PSF model, and there is no definite threshold for declaring a source ``resolved''.
We therefore perform S\'ersic profiles to our robust $\zspec>3.1$ sample and a sample of stars which are known point sources and thus unresolved.

We begin by selecting a sample of stars in the UDS field starting with the same criteria used in \citet{Weibel2024} but extending down to fainter magnitudes, $23 < m_{\rm f444w} < 27$, to better match the brightness of sources discussed in this study. 
To eliminate contaminants from this selection, typically compact galaxies at $1<z<4$, we require that stars satisfy two additional color cuts: $\textrm{F200W} - \textrm{F444W} < 0.75$ and $\textrm{F150W} - \textrm{F200W} < 0.1$.
Visual inspection of the images and SEDs confirms that these cuts yield a clean sample of stars. 
We ultimately identify 97 stars, to which we apply the same morphological fitting procedure and S/N cuts as used for our galaxy sample.

The posterior distributions of $r_{\rm eff}$ for the stellar sample are all bounded by the lower limit of the prior, with peaks in the distribution near 0.25\,px and an extended tail out to larger radii. 
To quantify the spread of these posteriors, we plot the stellar magnitude against the 95th percentile of the $r_{\rm eff}$ posterior, $r_\textrm{eff, 95\%}$, for each LW NIRCam filter in Figure~\ref{fig:morph}. 
For bright stars, we observe a floor in $r_{\rm eff, 95\%}$ around 1/3\,px, which we interpret as a systematic limit set by the PSF width and the accuracy of our PSF model. 
At magnitudes fainter than 25, $r_\textrm{eff, 95\%}$ increases, likely due to lower signal-to-noise ratios broadening the morphological posteriors. 
We observe that the slope of this increase is approximately 0.2\,mag per $\log(r/\textrm{px})$, consistent with a $\sqrt{\rm flux}$ dependence, confirming our expectation that the trend is driven by Poisson-dominated noise in the source centers.

We use the behavior of stellar magnitude versus $r_{\rm eff, 95\%}$ to determine whether sources in our sample are spatially resolved.
If a source falls within the locus defined by the stars, we classify it as unresolved. 
To assess this quantitatively, we fit a parametric model to the stellar distribution, described by a power-law with a plateau:
$$ \log_{10}\left(\frac{r_\mathrm{rsv}}{\mathrm{px}}\right) = \max\left(0.2 \cdot (\mathrm{mag} - 24.5) + b,\ a\right) \label{eqn:r_rsv}$$
where $a$ and $b$ are the parameters which are optimized separately for each filter. 
To ensure the curve fully encloses the stellar population, we compute the standard deviation of the residuals from the initial fit, shift both parameters upward by $+4\sigma_{\rm resid}$, and impose an upper limit of 1.5\,px., resulting in curve we denote as $r_{\rm rsv}(m)$.
The resulting curve, along with the best-fit $a$ and $b$ values for each filter, is shown in Figure~\ref{fig:morph}.

This analysis reveals the tradeoff in point-source selection: our conservative morphological cut, i.e.\ imposing a strict 1.5\,px upper limit, prioritizes sample accuracy at the potential cost of completeness.
While effectively minimizing contamination from compact galaxies, this approach may exclude genuine point sources below $\sim$26\,mag in F444W and $\sim$27.5\,mag in F277W/F356W, where the stellar locus continues to rise beyond our size cutoff (Figure~\ref{fig:morph}). We discuss the impact of this accuracy-completeness tradeoff on point-source selection in Section~\ref{sec:accuracy}.

For each source in our sample we select the LW filter based on the redshift to ensure that the filter probes the rest-optical continuum around $\sim$5500\AA, i.e.\ F277W at $3.1< \zspec < 4.6$, F356W at $4.6\leq \zspec < 6.1$, and F444W at $\zspec \geq 6.1$.
A galaxy of magnitude $m$ in the relevant filter is then classified as unresolved if \ $r_{\rm eff, 95\%} < r_{\rm rsv}(m)$.
After removing bad fits and objects which are too faint in the given LW band to confidently measure, we are able to classify the morphologies for 1305 (88\%) of our robust $\zspec > 3.1$ sample; of these 1199 (92\%) are identified as resolved, while the remaining 106 (8\%) are considered unresolved.

\begin{figure*}[ht!]
    \centering
    \includegraphics[width=\columnwidth]{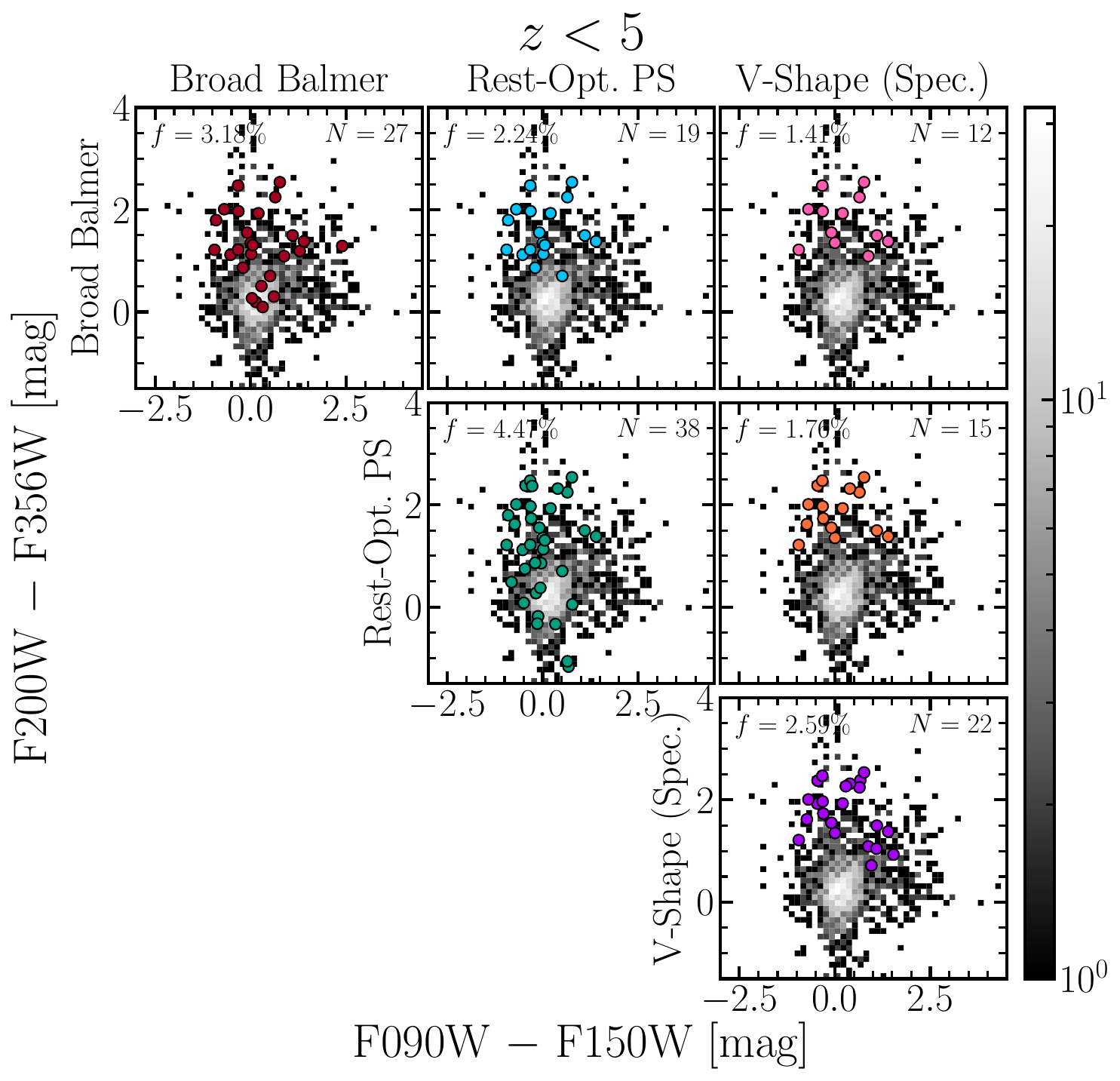}
    \includegraphics[width=\columnwidth]{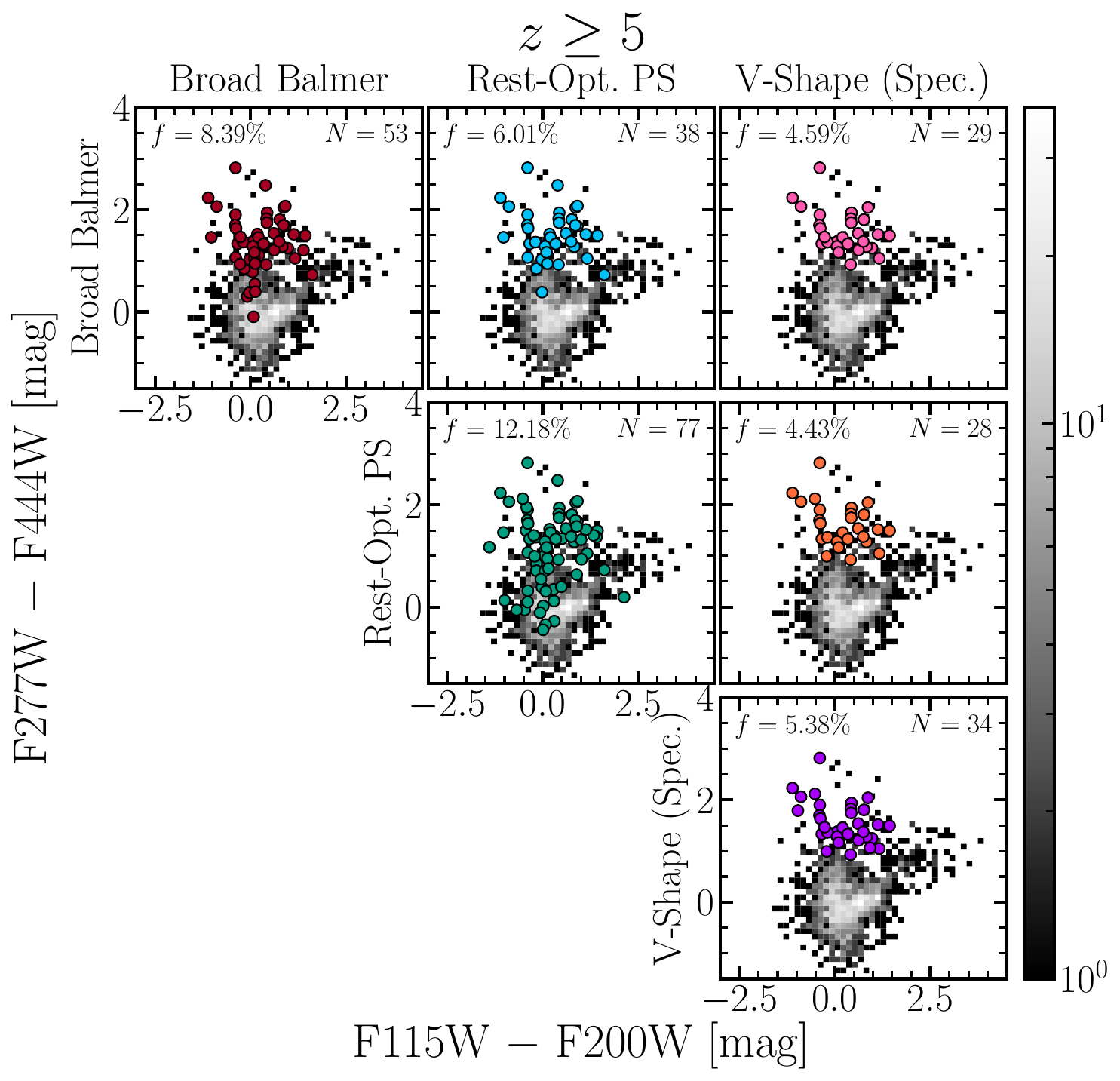}
    \includegraphics[width=\columnwidth]{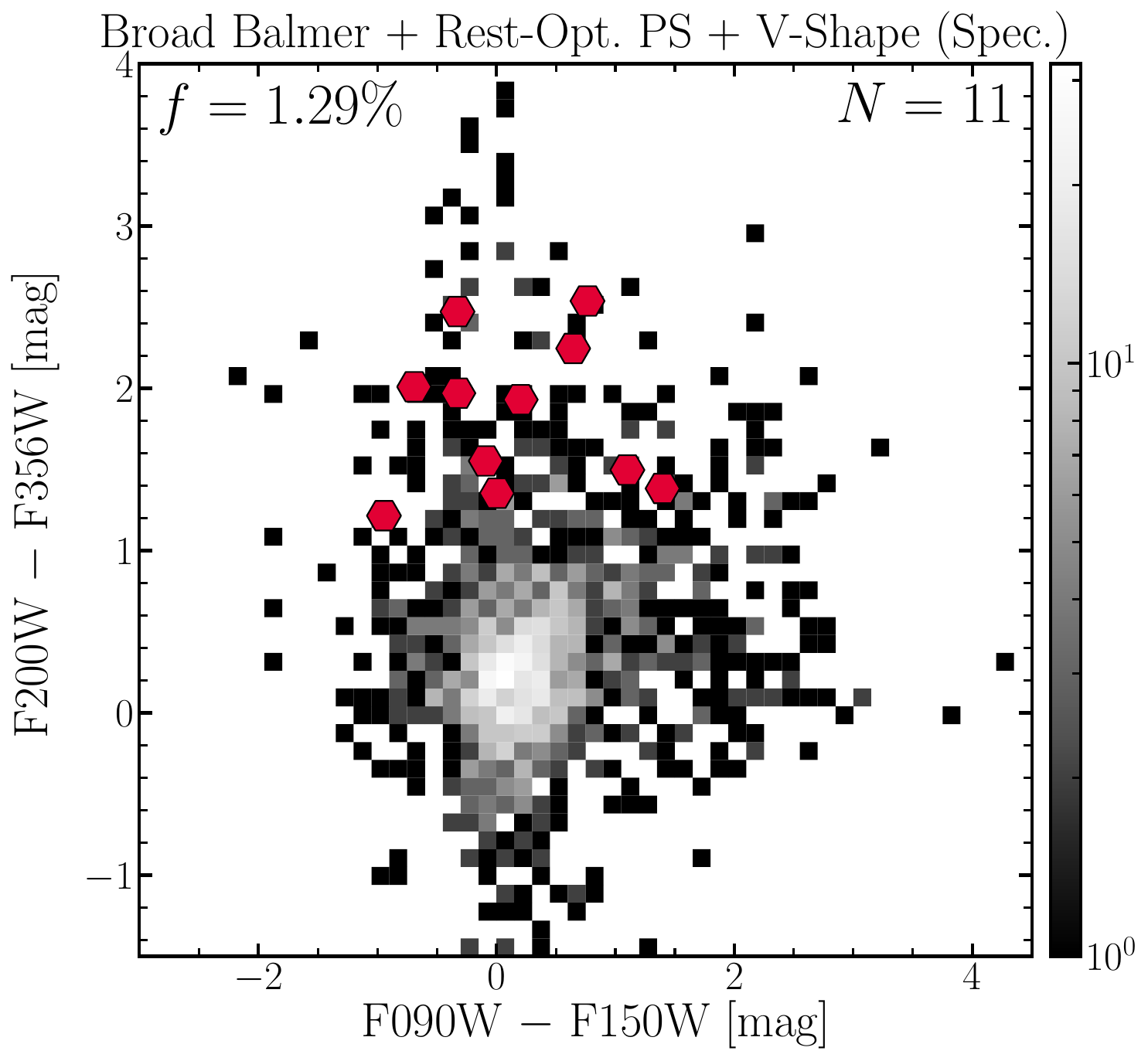}
    \includegraphics[width=\columnwidth]{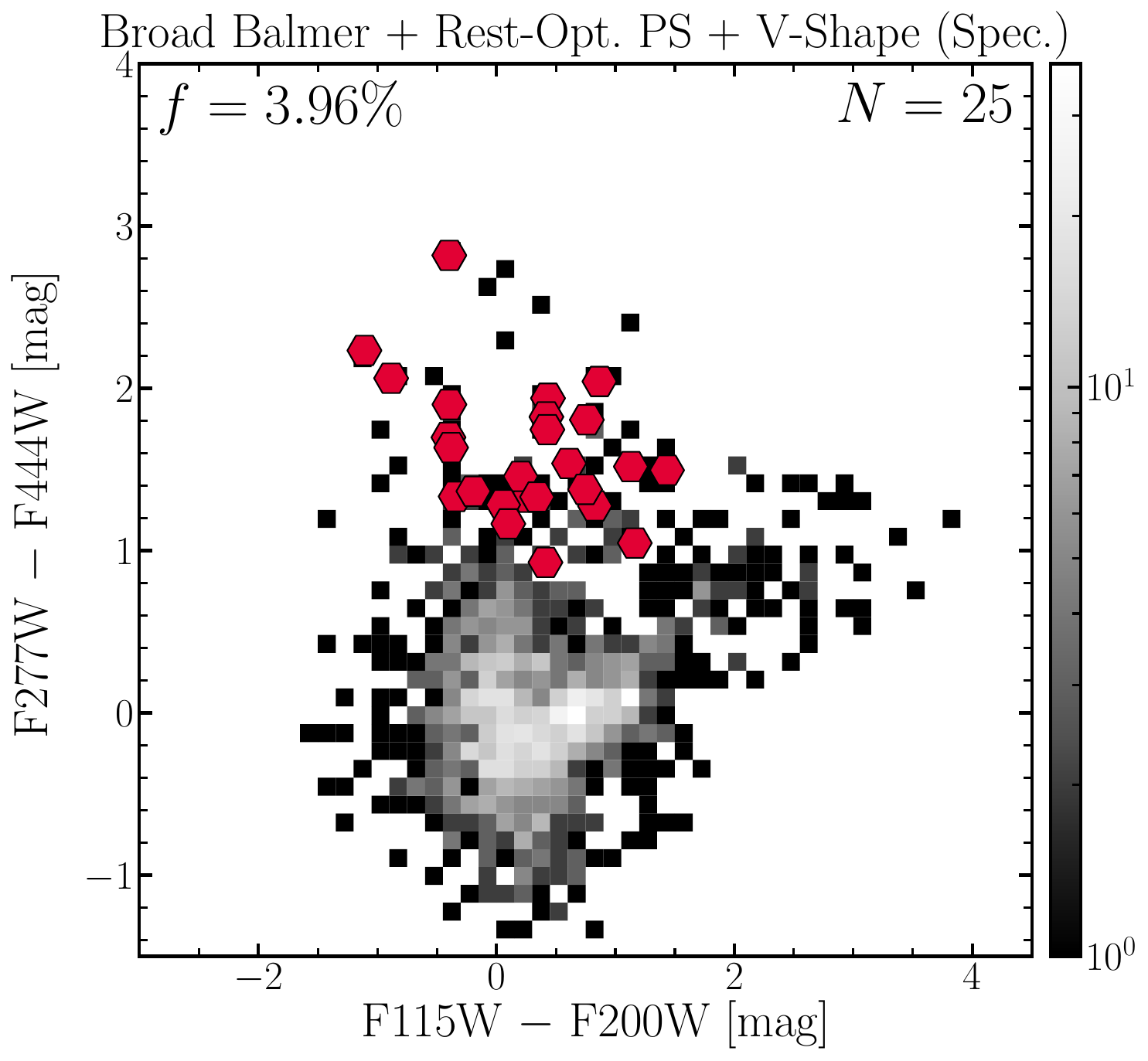}
    \caption{Color-color space distribution of RUBIES with robust $\zspec>3.1$ (grey histogram).
    We divide the sample into $\zspec < 5$ (right) and $\zspec > 5$ (left) and show the NIRCam F090W$-$F150W vs.\ F200W$-$F356W and F115W$-$F200W vs.\ F277W$-$F444W respectively which approximately probe rest-UV vs.\ rest-optical colors. 
    In the top row, we plot objects that satisfy our condition for broad-line, unresolved, and v-shaped features, along with their combinations, above the full distributions of the parent sample in each redshift regime.
    In the bottom row we plot the objects that satisfy all three criteria.
    Each panel includes the total number of objects shown and the fraction of the represented sample. Points are colored consistent with Figure~\ref{fig:euler}.}
    \label{fig:color}
\end{figure*}

\subsubsection{Dominant Point Source Component}\label{sec:res}

For sources classified as resolved in Section~\ref{sec:unr}, we investigate whether they still contain a dominant point-source component in their rest-optical emission. 
This analysis is particularly relevant for understanding systems where an unresolved nucleus might coexist with extended host emission. 
However, decomposing these components is inherently challenging due to model dependencies and prior choices. 
We therefore restrict this analysis to sources that already show either broad Balmer lines or v-shaped continua ($N=32$). 

We perform two-component modeling using \texttt{pysersic}, fitting each source with a combination of a point source and a S\'ersic profile. 
The central position of these two components are fixed. We follow much of the same procedure for the single S\'ersic profile fitting above but with an additional parameter $f_{\rm ps}$, the fraction of flux contained in the point source component. 
We assign $f_{\rm ps}$ a uniform prior; $\mathcal{U}(0,1)$.
We classify a source as having a dominant point source if the 95$^\textrm{th}$ percentile of the point-source flux fraction exceeds 50\% in the relevant LW filter. 
We find nine objects with a dominant rest-optical point source: RUBIES-EGS-15825, RUBIES-UDS-23438, RUBIES-UDS-24447, RUBIES-EGS-28812, RUBIES-UDS-33938, RUBIES-EGS-46724, RUBIES-UDS-57040, RUBIES-UDS-167741, and RUBIES-UDS-840721. 

\begin{figure*}[ht!]
    \centering
    \includegraphics[width=\columnwidth]{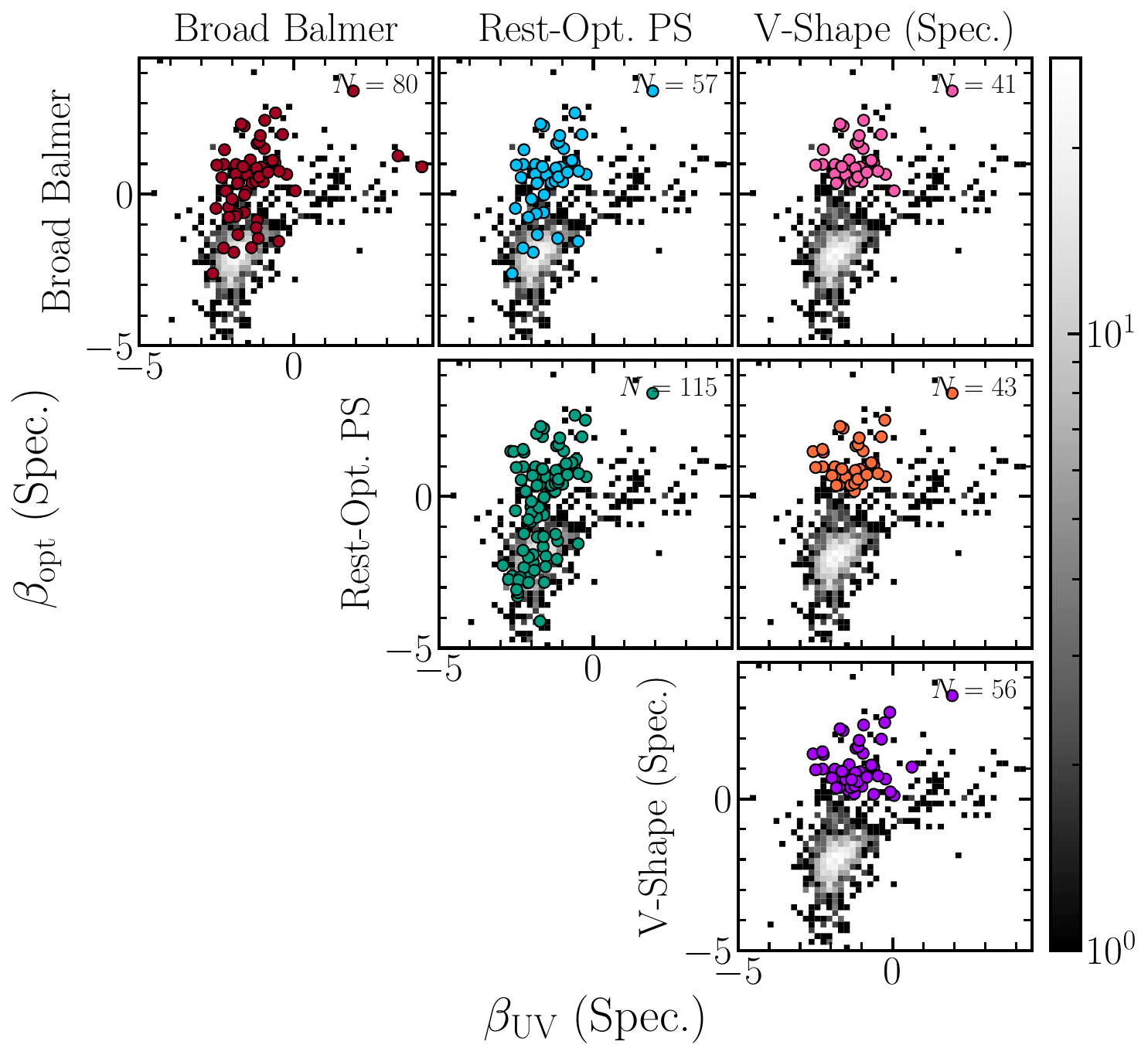}
    \includegraphics[width=\columnwidth]{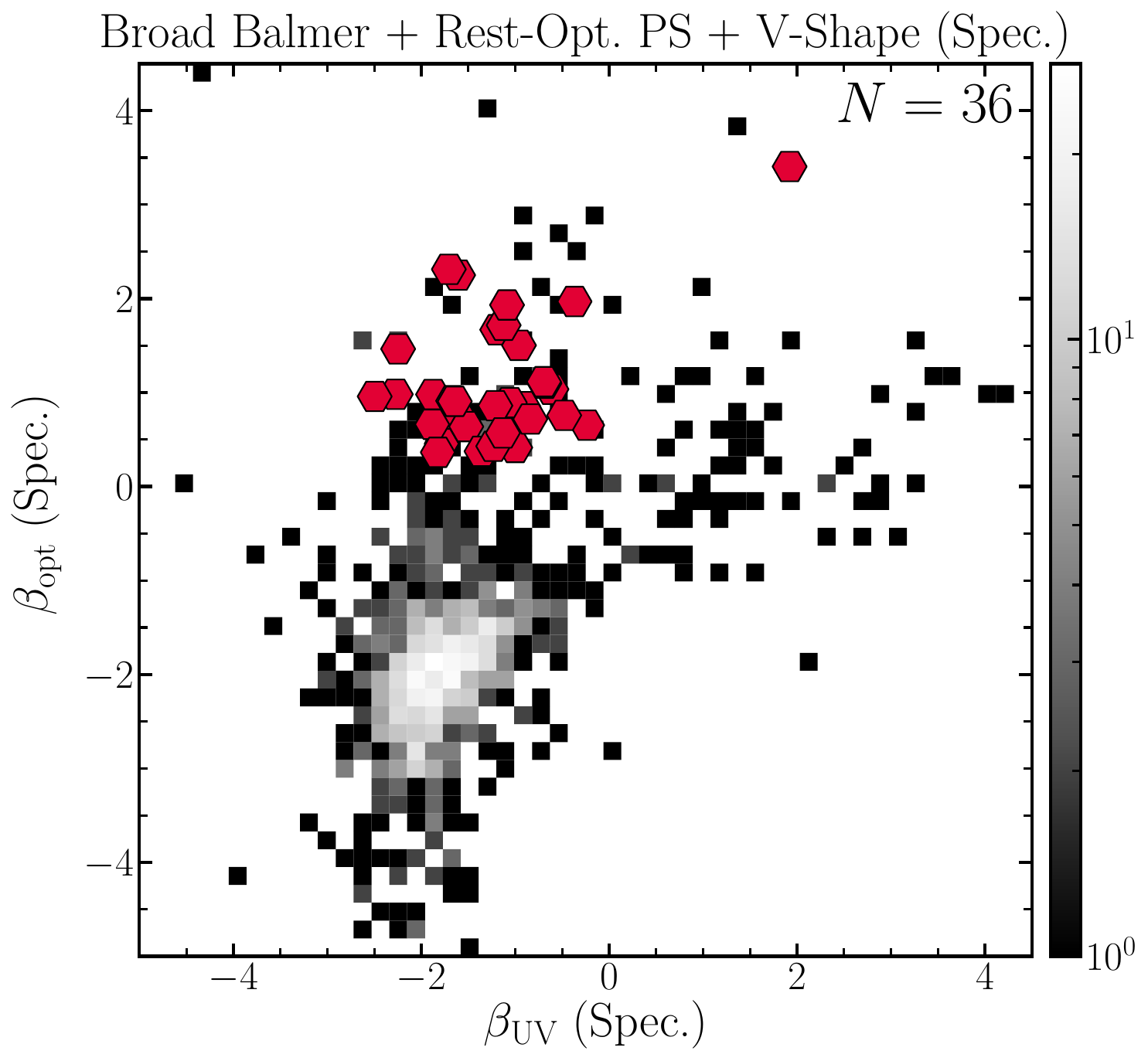}
    \caption{$\beta_\textrm{UV}$-$\beta_\textrm{opt}$ space distribution of RUBIES with robust $\zspec>3.1$ (grey histogram).
    On the right, we plot objects that satisfy our condition for broad-line, unresolved, and v-shaped features, along with their combinations, above the full distributions of the parent sample.
    On the left  we plot the objects that satisfy all three criteria. 
    Slopes are derived from PRISM spectroscopy and each panel includes the total number of objects plotted and their fraction of the represented sample.
    Points are colored consistent with Figure~\ref{fig:euler}.}
    \label{fig:beta}
\end{figure*}

\section{Spectroscopic LRD Selection} \label{sec:results}

Following Section~\ref{sec:methods}, we are able to make a confident measurement for all three key characteristics of LRDs in 1019 RUBIES galaxies, i.e.\ the majority (69\%) of our robust $\zspec > 3.1$ sample (85\% for the sample with PRISM spectroscopy).
We can therefore robustly assess, for the first time, how the combination of a v-shaped continuum, a dominant point-source in the rest-optical, and a broad-line detection correlate among the high-redshift galaxy population. 

\subsection{Colors and Intrinsic Spectral Slopes}\label{sec:colors_n_slopes}

In Figure~\ref{fig:color} we plot the color-color distribution of our $\zspec > 3.1$ sample and those that satisfy our condition for broad-line, unresolved, and v-shaped features. 
We divide the sample into two redshift bins, $z < 5$ and $z \geq 5$, and show NIRCam broad-band colors that approximately probe rest-UV vs.\ rest-optical colors in these regimes (F090W$-$F150W vs.\ F200W$-$F356W and F115W$-$F200W vs.\ F277W$-$F444W respectively). We note that the difference in sample size between the two redshift bins is, in part, driven by the fact that the \Oiii doublet is observed in the G395M grating for $z>5$, which enables broad line identification down to smaller widths. 
However, broadband colors can be significantly impacted by emission line contamination which depends on the spectroscopic redshift and the exact position of the emission line.
We therefore show the corresponding intrinsic spectral slopes, $\beta_\textrm{UV}$ vs.\ $\beta_\textrm{opt}$, as derived from spectroscopy in Figure~\ref{fig:beta} to compare and contrast the distributions.

\subsubsection{Broad Balmer Lines}

We begin by investigating the observed colors and intrinsic slopes of the full sample with robustly detected broad Balmer emission. 
Broad-line sources tend to be redder in the rest-optical than the full spectroscopic sample, partly due to luminous broad \Ha emission, which can bias rest-optical colors by up to 0.8\,mag.
In contrast, when comparing to spectral slopes, we find that the broad-line sample spans wider range of rest-optical slopes.
Their rest-UV slopes avoid the very reddest rest-UV tail, a region dominated by dusty star-forming galaxies.

\subsubsection{Rest-Optical Point Sources}

Galaxies that are unresolved or dominated by a point source at rest-optical wavelengths are distributed across much of color space. Again we see a preference for redder rest-optical colors compared to the full sample, although this may be in part due to the S/N criterion used to measure morphologies, i.e.\ that the rest-optical flux S/N$>$10 (see Section~\ref{sec:morph}). Similarly, when investigating spectral slopes, unresolved galaxies span the entirety of spectral-slope space populated by the sample, except for the reddest rest-UV slopes where the population is dominated by dusty star-forming galaxies that are typically larger in size. 

\subsubsection{Spectroscopic V-Shaped Continuum}

Enforcing a spectroscopic v-shaped selection will, by definition, restrict the sample to a limited area of spectral slope space. Similarly, v-shaped objects occupy a limited region of color space, but with slightly more scatter, attributable to emission line boosting, a moving break location that depends on the true source redshift, and low S/N in the bluest filters for the reddest sources. Roughly four-fifths of the v-shaped sources are unresolved. 
The remaining fifth may be related to the population of dusty starbursts with v-shaped continua and ALMA 1.2\,mm detections from \citet{Labbe2023b}. 
Many of these galaxies in our sample, while resolved, are compact and show hallmarks of star formation, including strong but narrow \Ha and \Nii, and may be related to other populations of compact, red, dusty, star-forming galaxies studied with JWST \citep[e.g.][]{akinsTwoMassiveCompact2023,Williams2024,Perez2024,Barro2024}.

\begin{figure*}[ht!]
    \centering
    \includegraphics[width=0.95\columnwidth]{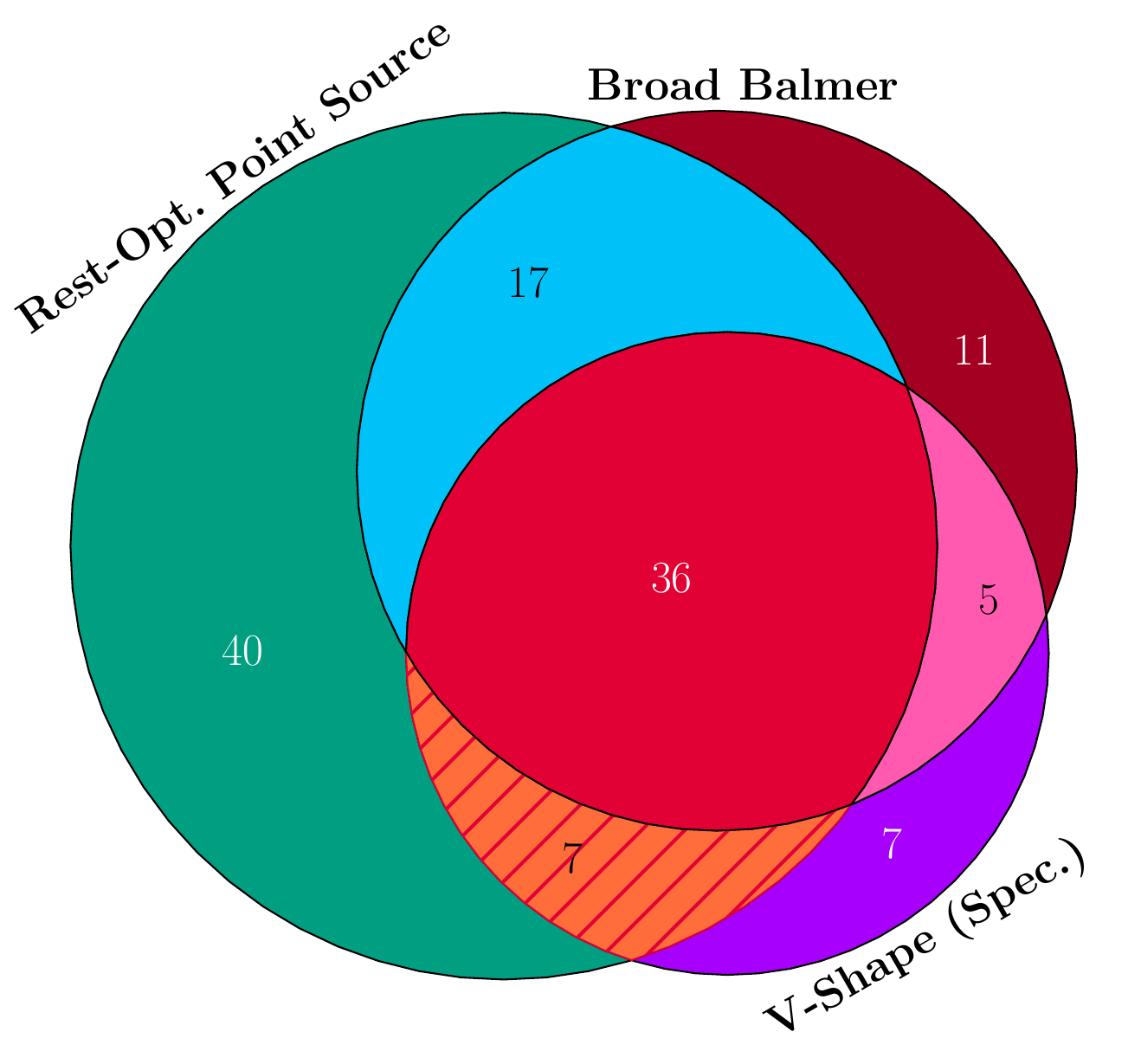}
    \includegraphics[width=1.05\columnwidth]{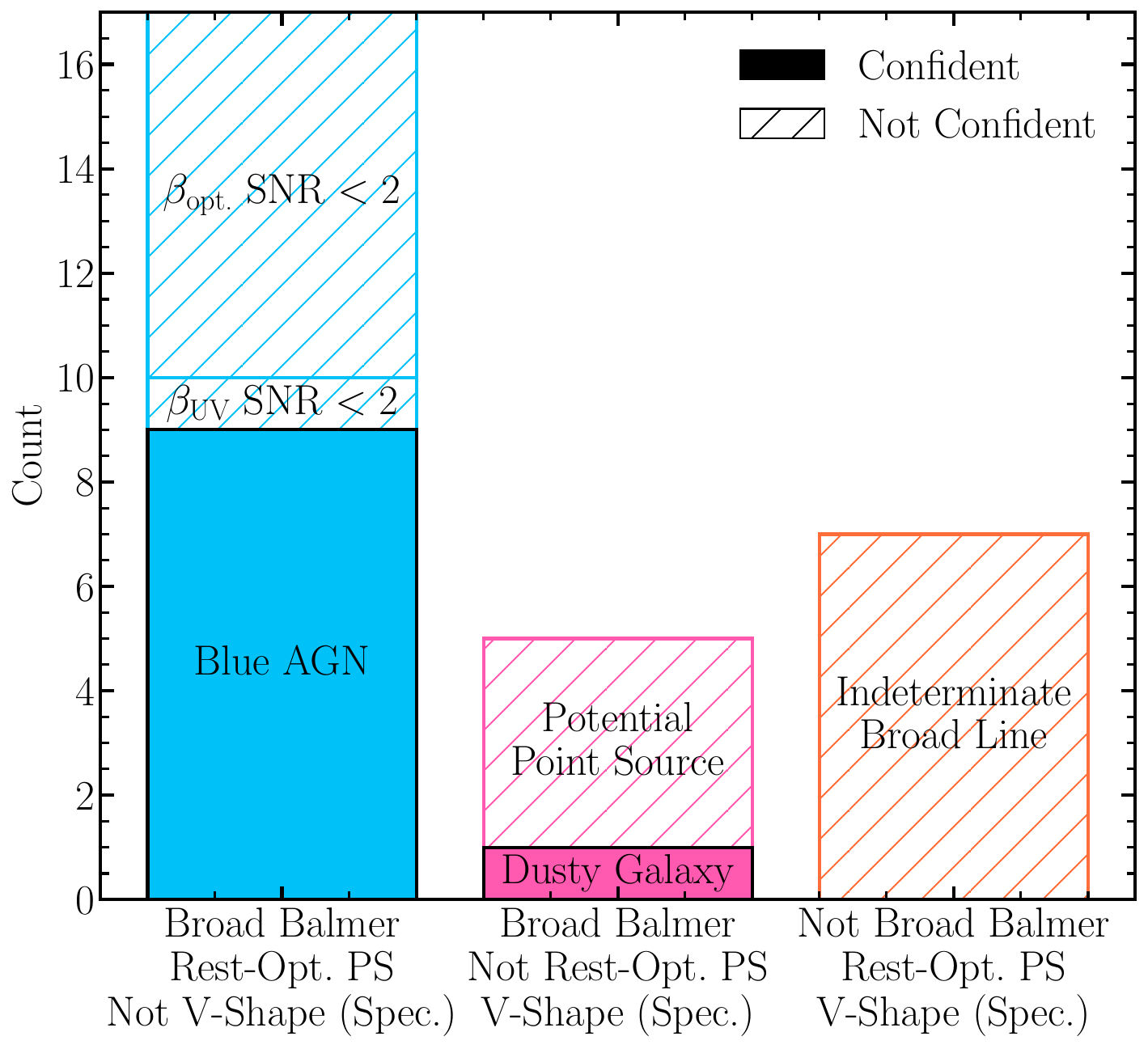}
    \caption{Left: Euler diagram displaying the overlap between the three LRD characteristic criteria: a broad Balmer line, a dominant point source component at rest-optical wavelengths, and a spectroscopic v-shaped continuum. Right: Histogram showing the composition of overlaps in the Euler diagram where only two of the three features are measured. 
    We note that objects with a rest-optical point source and a v-shaped continuum are 80\% likely to have a broad line. 
    In fact, in the remaining 20\% we cannot reject a broad line, but our quality cuts restrict us from being fully confident.
    This suggests an underpinning physical link between the three LRD features.}
    \label{fig:euler}
\end{figure*}

\subsection{Relationship Between Typical LRD Features}\label{sec:lrd}

We investigate how the full sample of galaxies with broad Balmer lines maps onto the v-shape and morphology selections. 
We find that the broad-line population divides into roughly three groups:
\begin{enumerate}
    \item Resolved systems with broad lines. These span a variety of intrinsic slopes and likely represent galaxies with a non-dominant AGN.
    \item Unresolved systems with broad lines having blue rest-optical and rest-UV slopes that are likely comprised of typical AGN-dominated systems.
    \item Broad-line systems with v-shaped continua that are spatially unresolved. 
\end{enumerate}

Figure~\ref{fig:euler} presents the Euler diagram exploring the relationship between all three features. 
Most remarkably, we find that if one starts with all sources having a v-shaped continuum, then either an unresolved rest-optical morphology or a broad-line selection will yield predominantly objects where \emph{all three} features are robustly detected ($>80\%$; red intersection). The remaining $\sim20\%$ that do not satisfy the third criterion are overwhelmingly candidate LRDs rather than interlopers from distinctly different populations. 

Specifically, while seven of the v-shaped rest-optical point sources (i.e.\ the orange intersection) technically lack a broad line detection, they are all classified as indeterminate broad lines, i.e.\ all are cases with data limitations where the presence of a broad line is ambiguous due to a DQ issue or a lack of coverage of the forbidden lines.
Similarly, of the v-shaped with broad-line sources (pink intersection), we find they are all well-described by an additional point-source component, but that the prominence of this component falls below the strict limit ($>50\%$) we require in this work. We note that one out of five sources, RUBIES-UDS-5496, shows strong \Nii emission not seen in any of our spectroscopic LRDs, as defined in Section~\ref{sec:lrd}. 
We assert that this is likely a dusty, star-forming galaxy hosting an AGN and not consistent with the population of spectroscopic LRDs explored in this work.  

\subsection{A Spectroscopic Definition of LRDs}

We find that all intrinsically v-shaped spectra with dominant point sources in rest-optical imaging exhibit broad Balmer lines, provided the data quality permits their identification.
This conclusion arises from a systematic analysis across galaxy color, morphology, and spectral shape, made possible by the RUBIES selection.
Moving forward, we define a spectroscopic LRD as a source that simultaneously satisfies these three criteria: a broad Balmer line, a v-shaped continuum, and a dominant point source component in rest-optical imaging.
Applying these criteria, we identify 36 spectroscopic LRDs in the RUBIES dataset, the largest such spectroscopic sample to date.  
NIRCam imaging, PRISM and G395M spectra, and broad line fits for these sources are provided in Appendix~\ref{app:lrd}.
In addition, we identify seven v-shaped, rest-optical point sources for which broad lines could not be definitively confirmed.
Although we consider it plausible that these sources are also spectroscopic LRDs, we do not include them in our spectroscopic LRD sample as DQ issues prevent the definitive measure of a broad Balmer line. 
These objects are nonetheless presented in Appendix~\ref{app:lrd} for completeness.

\begin{figure*}[ht!]
    \centering
    \includegraphics[width=\columnwidth]{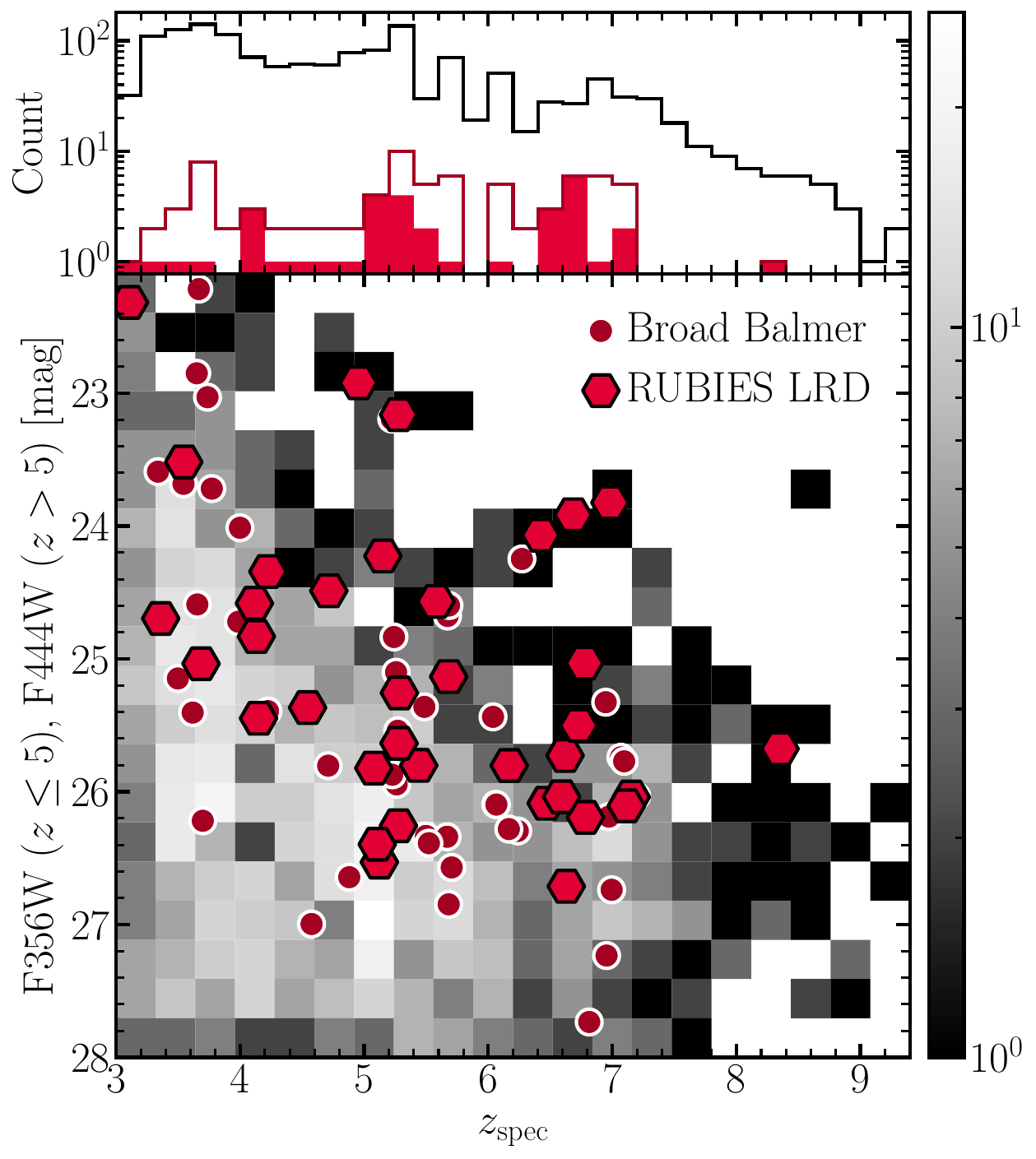}
    \includegraphics[width=\columnwidth]{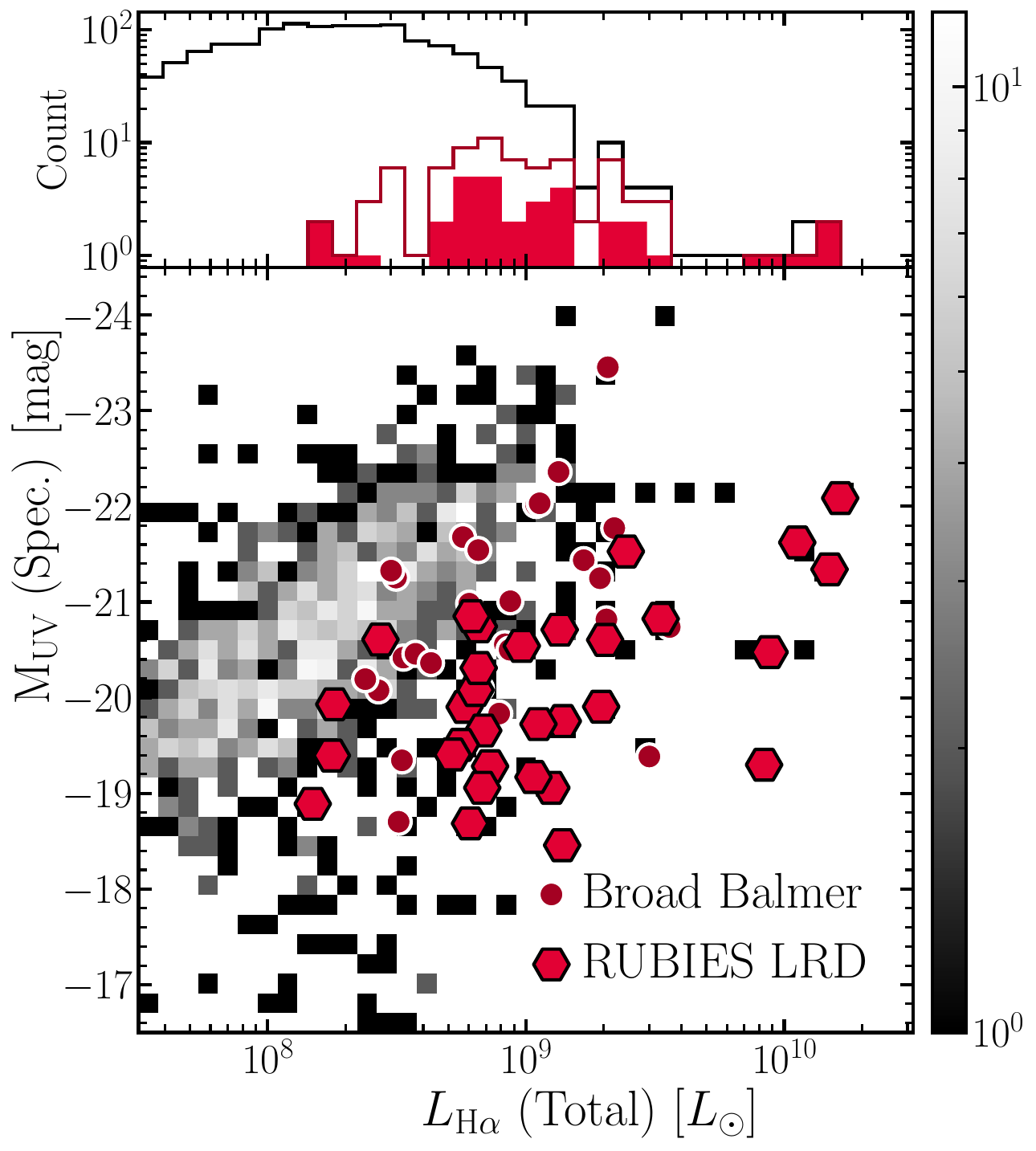}
    \caption{Redshift vs. NIRCam F356W (for $\zspec \leq 5$) or F444W (for $\zspec > 5$) is shown on the left, and absolute UV magnitude ($\rm M_{UV}$) at 1500\AA, derived from rest-UV spectroscopic slopes, is plotted against total \Ha luminosity on the right. 
    RUBIES with robust $\zspec > 3.1$ are shown as a grey histogram. 
    Broad-line sources are indicated with dark red circles, and LRDs, defined as galaxies with v-shaped continua, broad Balmer lines, and dominant rest-optical point sources, are shown as light red hexagons.
    At fixed $L_{\rm H\alpha}$ LRDs are faint in the rest-UV compared to the full population.
    Conversely, extreme \Ha emitters ($L_{\rm H\alpha}\gtrsim10^{10}\,\Msun$) are dominated by LRDs, i.e.\  LRDs constitute the most luminous \Ha emitters at fixed UV luminosity.
    \label{fig:muvha}
    }
\end{figure*}

We briefly investigate how the properties of spectroscopic LRDs compare to the broad Balmer line sources and the robust $\zspec > 3.1$ RUBIES sample.
The left panel of Figure~\ref{fig:muvha} demonstrates a wide redshift distribution for the broad-line objects, with the LRDs spanning a similar range in both redshift and rest-optical magnitude. 
However, it is clear that LRDs are comprised of a distinct family of properties.
Typical AGN, with broad lines and blue power-law continua, separate themselves primarily in continuum shape from the LRDs recovered in this work.

To further highlight their discrepancies, we compute UV luminosities at 1500\AA\ using the rest-UV continuum fits to the spectra from Section~\ref{sec:cont}. 
In the right panel of Figure~\ref{fig:muvha} we show the total \Ha luminosity vs. M$_\textrm{UV}$ for the full robust $\zspec>3.1$ sample, the broad Balmer line sources, and the LRD sample. 
Although both the broad Balmer line sources and LRDs are offset from the bulk of the population, LRDs are the most \Ha luminous sources at all $M_{\rm UV}$.
However, the LRDs also show a wide diversity of UV-to-optical ratios, exceeding the much narrower range in $L_{\rm H\alpha}$ at fixed $M_{\rm UV}$ spanned by the ``typical'' broad-line sources. 
We defer a full exploration of the interconnectedness of LRD properties and a physical interpretation to future work.

Overall, we conclude that LRDs -- when selected using spectroscopy -- comprise a distinct family of properties. 
Standard broad Balmer sources with power-law continua are well-represented in our sample, but are distinct in both continuum shape and \Ha/UV luminosity from the spectroscopic LRDs. 

\section{Photometric LRD Selection} \label{sec:phot_sel}

Using an empirical approach to the RUBIES spectroscopic dataset, we have identified a population of LRDs that have broad Balmer emission lines, v-shaped continua, and compact rest-optical morphologies. 
This is in contrast to typical LRD searches that, to date, have often been based on photometry alone \citep[e.g.][]{Barro2024,Labbe2023b,Kocevski2024,Kokorev2024,Akins2024}. 
This raises the major question of whether photometric selection of LRDs yields the same sources as our spectroscopic selection. 
\citet{Greene2024} showed that $\approx$80\% of photometric LRDs from the selection of \citet{Labbe2023b} indeed have broad lines and v-shaped continua, albeit from a small sample of 12 sources. 

In this section, we use our large spectroscopic sample of LRDs to evaluate the accuracy and completeness rates of popular photometric LRD selection strategies. 
We then investigate both contaminants and missed LRDs in the photometric samples.

\subsection{Broad-line and LRD Success Rates}\label{sec:bl_success}

Two primary methods have been used to select LRD candidates in photometric surveys.
The first is based on identification of v-shaped broad-band photometric SEDs, and the second on the selection of red broad-band rest-optical colors, sometimes additionally requiring blue rest-UV colors. 
All methods impose an additional compactness criterion. 
These different approaches have also been applied in the RUBIES parent fields, the EGS and UDS, by two large photometric searches, \citet{Kocevski2024} using v-shape selection and \citet{Kokorev2024} using multi-color selection, providing an ideal opportunity to assess their effectiveness. 
To quantify the success rates of these photometric methods we cross-match our spectroscopic sample with the public photometric catalogs from each study. 

\begin{table*}
\renewcommand{\arraystretch}{1.2}
\caption{Photometric LRD Selection Comparison}
\begin{center}
 \begin{tabular}{lccc}\hline\hline
    & Kocevski et al. (2024) & Kokorev et al. (2024) & $\rm F277W-F444W>1.5$ \\ 
   Photometric Constraint & Photometric V-Shape & Multi-Color & Single-Color \\
   Morphological Constraint & Compact & Compact & Rest-Opt. Point Source \\ \hline
   RUBIES Matches &  53/341 (15.54\%)  & 40\textsuperscript{a}/259 (15.44\%) & 29 \\   
  Robust $\zspec>3.1$ (\& PRISM)   &  47 (45)  & 32 (28) & 21 (20) \\   Brown Dwarfs & 0 & 1 & 0 \\ \hline
  & \multicolumn{3}{c}{Robust $\zspec>3.1$}\\ \hline
  Confident Broad Balmer & 31/47 (65.96\%) & 26/32 (81.25\%) & 16/21 (76.19\%) \\  
  Unknown Broad Balmer (F444W $>$ 26.5) & 10/47 (21.28\%) &  1/32 (3.12\%) & 3/21 (14.29\%)  \\
  Unknown Broad Balmer (F444W $<$ 26.5)\textsuperscript{b} &  5/47 (10.64\%) & 5/32 (15.62\%)& 2/21 (9.52\%) \\\hline
  & \multicolumn{3}{c}{Photometric Sample Makeup}\\\hline
  $\zspec > 3.1$ \& PRISM \& $\rm F444W < 26.5$ & 33 & 27 & 18 \\
  Uncertain & 11/33 (33.33\%) & 8/27 (29.63\%) &  5/18 (27.78\%) \\
  Confident Contamination & 1/33 (3.03\%) & 2/27 (7.41\%) & 1/18 (5.56\%) \\
  RUBIES LRD &  21/33 (63.64\%) & 17/27 (62.96\%)  & 12/18 (66.67\%) \\
  \hline\hline
  RUBIES LRD Accuracy & 21/22 (95.45\%) & 17/19 (89.47\%) & 12/13 (92.31\%) \\ 
  RUBIES LRD Completeness &  21/34 (61.76\%) &  17/34 (50.00\%) & 12/34 (35.29\%) \\ 
  \hline\hline
 \end{tabular}
\tablefoot{
\textsuperscript{a}Although there are 41 matches to \citet{Kokorev2024}, we find that IDs 1470 and 59971 are the same object in that catalog. This object is only counted once in our analysis.\\
\textsuperscript{b}All uncertain broad Balmer lines with $\rm F444W<26.5$ are indeterminate broad, i.e.\ lack \Oiii or H$\alpha$ in G395M, or suffer from a DQ issue.
}
\label{tab:phot_samples}
\end{center}
\end{table*}

We begin with the selection of \citet{Kocevski2024}, which performed double power-law fitting to HST and NIRCam photometry to select v-shaped SEDs. 
The exact filters used depend on the photometric redshift, but typically comprise three broad-band filters both blue- and redward of the estimated Balmer limit. 
We identify 53 sources with matches to the RUBIES spectroscopy, 47 of which have robust $\zspec > 3.1$. 
The majority (65\%) of these sources show broad Balmer emission lines, in line with an early estimate of the broad line fraction by \citet{Kocevski2024} that was based on a small fraction of the RUBIES dataset. 
Of the remaining 35\%, two thirds are fainter sources where the presence of a broad line cannot be ruled out with the depth of the existing data, while the remainder suffer from data quality issues or lack G395M coverage of forbidden or Balmer lines.

This photometric sample spans a wide range in F444W magnitudes, and nearly half the sample is fainter than the RUBIES LRDs. 
For a representative analysis of the spectroscopic LRD recovery, we restrict our further comparison to sources brighter than $\rm F444W<26.5$. 
This limit is chosen based on the depth of the RUBIES PRISM spectra, corresponding to a median S/N $\sim$ 3 per resolution element for a well-centered point source \citep[see][]{deGraaff2024d}. 
Moreover, as described in Section~\ref{sec:unr} and shown in Figure~\ref{fig:morph}, the distinction between genuine point sources and compact galaxies becomes increasingly ambiguous at these fainter magnitudes. 
We find that 21 of the total 34 spectroscopic LRDs satisfying this magnitude criterion are recovered by the selection of \citet{Kocevski2024}. 
This translates to a success rate, i.e.\ a spectroscopic LRD  completeness, of approximately 60\%. 
We summarize these results in Table~\ref{tab:phot_samples}. 

Next, we turn to the color-selected sample of \citet{Kokorev2024} which requires two red broad-band colors at $\sim 2-4\,\micron$ as well as a single blue broad-band color at $<2\micron$. 
Although photometric redshifts are not explicitly used, in practice these criteria translate to, and were optimized for, selection on rest-UV and rest-optical colors of high-redshift sources. We find 40 unique cross-matched sources, with 32 having robust $\zspec > 3.1$.
This sample contains a remarkably high fraction of broad-line sources, 81\%, corroborating the earlier finding of \citet{Greene2024} who evaluated the multi-color selection of \citet{Labbe2023b} that is  similar to that of \citet{Kokorev2024}. T
he majority of the remaining 19\% have indeterminate broad Balmer lines, i.e.\ lacking G395M coverage of either \Oiii or \Ha, or suffer from DQ issues, and the true broad line fraction may therefore be even higher. 
We find that 17 of the 34 RUBIES LRDs at $\rm F444W<26.5$ are recovered, which implies a spectroscopic LRD completeness of 50\%.

Finally, several studies have proposed the use of a single, extremely red rest-optical color to select LRDs \citep[e.g.][]{Akins2024,Barro2024,Greene2024}. 
Multi-color selection has clear advantages, but it relies on the availability of a large number of photometric filters, which do not exist for the widest-area JWST programs such as COSMOS-Web or pure-parallel programs \citep[e.g.][]{Casey2023,Williams2025:panoramic}. 
Although not used before in the RUBIES fields, we also assess the use of the $\rm F277W-F444W$ color paired with our requirement of a dominant rest-optical point source (see Section~\ref{sec:morph}). 
Following \citet{Akins2024}, we first impose $\rm F277W-F444W>1.5$;  
due to the restrictive nature of this requirement it selects far fewer galaxies (29, of which 21 have robust $\zspec>3.1$). 
The broad line fraction among this sample is very high (76\%, see Table~\ref{tab:phot_samples}), however, the recovery of spectroscopic LRDs is highly incomplete (35\%) and the single-color selection thus performs significantly worse than the other photometric selections. 
If the color criterion is instead relaxed to $\rm F277W-F444W>1.0$ the LRD completeness improves (65\%), but at the expense of the recovered broad line fraction, which drops to 61\%.

\begin{figure*}[ht!]
    \centering
    \includegraphics[width=0.480\textwidth]{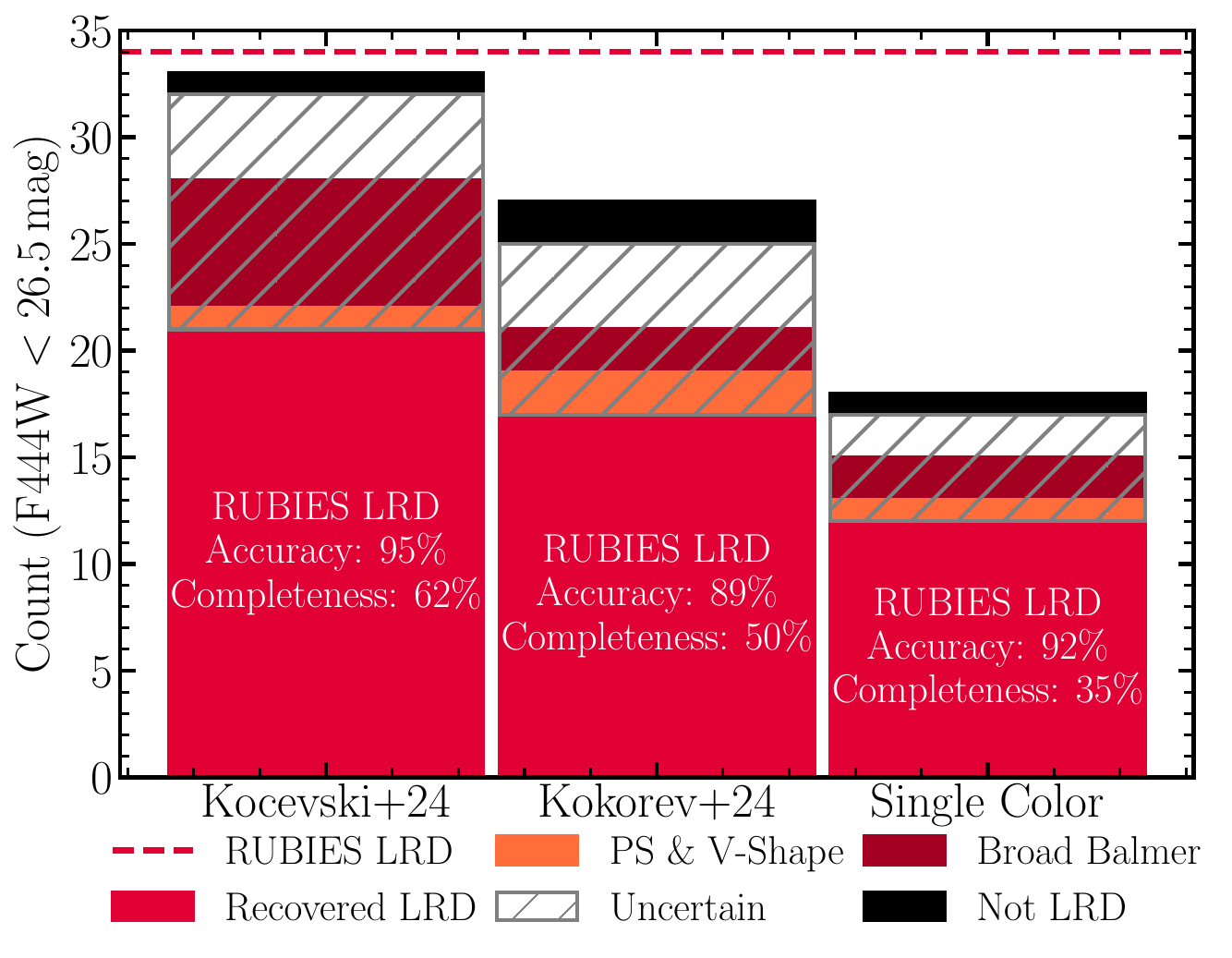}
    \includegraphics[width=0.512\textwidth]{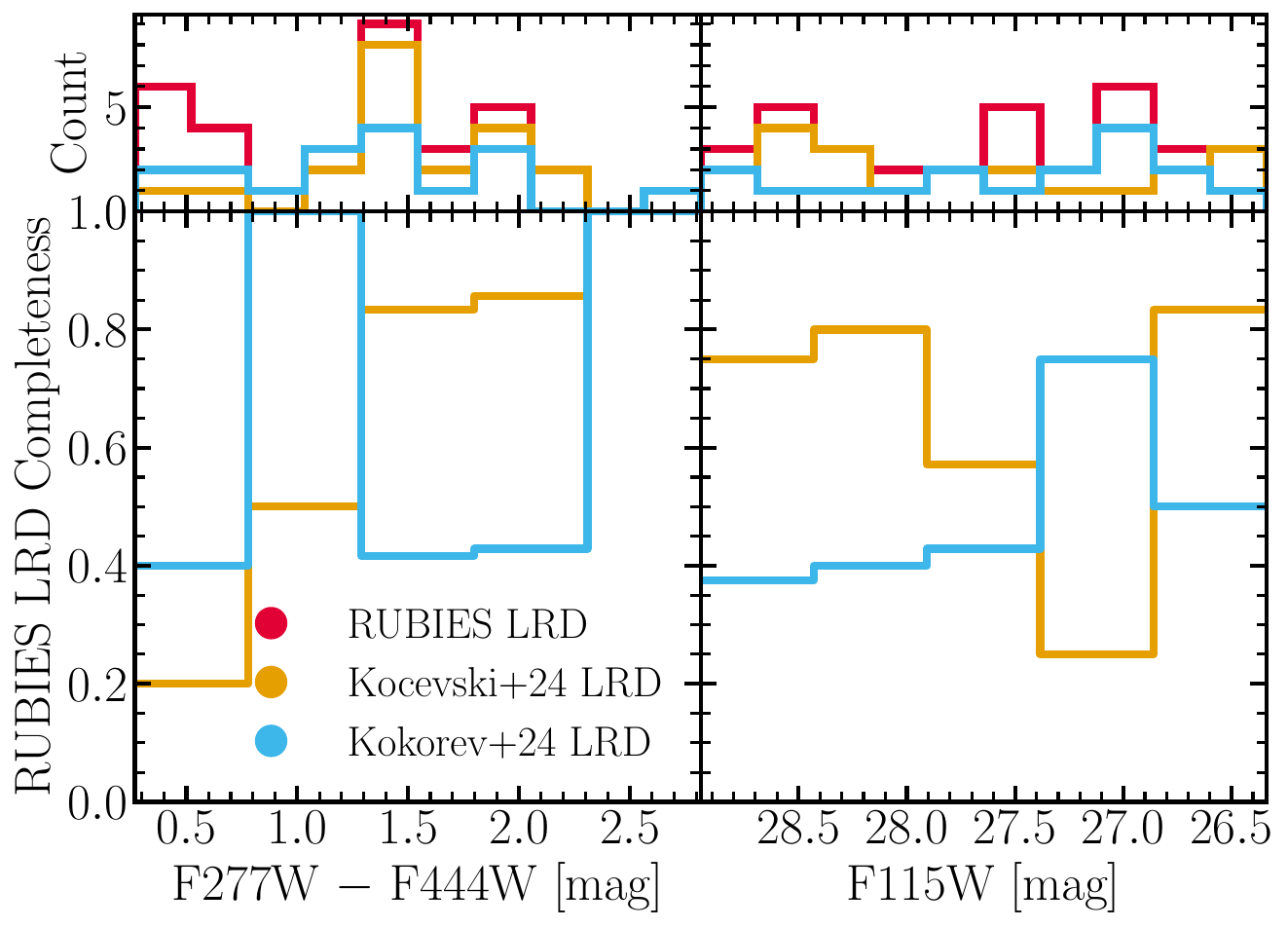}
    \caption{
    The breakdown of photometric LRD selection methods at $\rm F444W < 26.5$. 
    The left panel shows how photometric LRDs are distributed among spectroscopically classified categories: spectroscopic LRDs (red), rest-optical point sources and v-shaped galaxies (orange), broad Balmer line sources (red), non-LRDs (black), and galaxies with unreliable feature measurements (grey hatched). 
    Overall, photometric LRD selections are accurate but incomplete relative to spectroscopic identifications.
    The right panel investigates the completeness of the RUBIES photometric LRDs as a function of F277W$-$F444W color (left) and F115W magnitude (right). 
    The top plots show histograms of RUBIES LRDs (red), matched LRDs from \citet[orange]{Kocevski2024}, and \citet[blue]{Kokorev2024}.
    The bottom plots display the binned completeness of each selection method relative to the RUBIES sample. 
    We find that \citet{Kocevski2024} is incomplete at bluer colors, while \citet{Kokorev2024} is primarily incomplete at fainter rest-UV magnitudes.}
    \label{fig:phot-hist}
\end{figure*}

\subsection{Accuracy and Contamination}\label{sec:accuracy}

The different photometric LRD selections are all successful in selecting sources with broad Balmer lines (upwards of 65-80\%), and select a large number of spectroscopic LRDs. 
This is perhaps not surprising in the context of our results of Section~\ref{sec:results}, where we demonstrated that a compact rest-optical morphology and v-shaped continuum are highly predictive of the presence of a broad line. 
However, unlike the spectroscopic v-shape measurements, the photometric colors and v-shapes can be biased by strong emission lines that substantially boost even broad-band filters. 
So far, we have only focused on the success rates of broad line and LRD recovery. 
We now turn to the contaminants in these samples and the accuracy of the photometric selections.

To investigate the possible contaminants in the photometric samples, we first restrict our analysis to the magnitude-limited sample ($\rm F444W<26.5$). 
Whereas initial follow-up studies of LRDs highlighted brown dwarf stars in the Milky Way as possible interlopers \citep[e.g.][]{Langeroodi2023,Burgasser2024,Greene2024}, we find no such contaminants in the photometric samples, demonstrating the success of the color cuts that were imposed to filter out cool stars. 
In fact, even when expanding to the full sample we only find one faint brown dwarf in the sample selected by \citet{Kokorev2024}. 
The other likely class of contaminants consists of AGN or star-forming galaxies with high equivalent width emission lines, which bias the observed rest-optical colors measured from photometry, as demonstrated by, e.g.\ \citet{Kocevski2023,Hainline2025}. 
However, we only identify 1-2 such sources in the magnitude-limited photometric samples, which are broad-line AGN with blue spectroscopic continua but very strong emission lines. 

The sources for which we can confidently conclude that they are LRDs or contaminants therefore account for $\sim65\%$ and $\sim5\%$, respectively, of the magnitude-limited photometric selections (see Table~\ref{tab:phot_samples} and left panel of Figure~\ref{fig:phot-hist}). 
This still leaves a substantial fraction of sources of uncertain nature, which can be in equal parts ascribed to data quality issues preventing a robust broad-line measurement (e.g. chip gaps, lack of forbidden lines), and low continuum S/N limiting a robust conclusion on whether the continuum is indeed v-shaped. 
To define the \emph{accuracy} of the photometric LRD selections we therefore only consider the sources for which we can robustly determine both the broad line and continuum shape criteria. 
This spectroscopic LRD accuracy is very high ($\sim90-95\%$) for all three photometric selections (as discussed in Section~\ref{sec:bl_success}) and would still be high ($\sim80\%$) even in the extreme case that all uncertain sources without data quality issues are considered contamination.

Although the accuracy of the photometric LRD selection is very high, we emphasize that it only applies to the magnitude-limited sample of $\rm F444W<26.5$, and it is as of yet unclear how this would extend to the fainter LRD candidates. 
For these fainter candidates we may expect low-mass star-forming galaxies, i.e. compact sources with strong emission lines, to form a more prominent source of contamination. 
This is difficult to confirm or rule out with present data, as deep low- and medium-resolution spectra are needed to determine the broad-line and continuum properties. 
Perhaps even more challenging is the fact that lower-mass galaxies are increasingly compact and therefore difficult to distinguish from true point sources. 
This effect is especially pronounced in the F444W filter where, at the available depths in the UDS and EGS, the morphological classification must either become inaccurate or incomplete at $\rm F444W > 26.5$ based on our fits to foreground stars.

\subsection{Incompleteness}\label{sec:incompleteness}

Most surprising is the fact that the photometric LRD selections are only able to recover up to 60\% of the spectroscopic LRDs, even when restricting to $\rm F444W<26.5$. 
This high incompleteness impacts, for example, the inferred number densities and luminosity functions \citep[e.g.][]{Kocevski2024,Kokorev2024}, and further increases the suggested tension between the inferred black hole properties and theoretical galaxy formation models \citep[e.g.][]{Habouzit2025}. 

We investigate the origin of this incompleteness in the photometric samples in the right panel of Figure~\ref{fig:phot-hist}, which shows the spectroscopic LRD completeness as a function of the broad-band color F277W$-$F444W, a proxy for the rest-optical slope, and the F115W magnitude, tracing the rest-UV brightness. 
This enables us to identify the trade-offs made in each selection method and its impact on LRD completeness.
The photometric v-shape selection of \citet{Kocevski2024} requires a strongly rising red rest-optical continuum, and additionally filters out sources that are likely to have strong emission lines based on the F277W$-$F356W and F277W$-$F410M colors. 
As a result, we find that the photometric v-shape selection of \citet{Kocevski2024} becomes increasingly incomplete toward bluer rest-optical colors. 
Relaxing the rest-optical color criteria would likely improve the completeness, but also would allow more extreme emission line galaxies with intrinsically blue continua to enter the sample. 
The multi-color selection of \citet{Kokorev2024} allows for more modest rest-optical colors, and the completeness is therefore approximately even across a wide range F277W$-$F444W color. However, \citet{Kokorev2024} imposed strict S/N criteria on the rest-UV ($\sim1-2\,\micron$) magnitudes, which translates into an incompleteness predominantly at fainter rest-UV magnitudes. Lowering this rest-UV brightness threshold would likely increase the risk of contamination from red interlopers, such as brown dwarfs and compact dusty star-forming galaxies at lower redshifts. 

Because these selection methods are incomplete in different regimes, it raises the question whether their combination would yield an improved sample. 
Indeed, we find only a modest overlap of 11 RUBIES LRDs between the \citet{Kocevski2024} and \citet{Kokorev2024} samples. 
The combination of the two samples therefore recovers 27 out of 34 spectroscopic LRDs in the magnitude limited sample, corresponding to a joint completeness of 79\%. 
This indicates that while neither method is individually complete, they are highly complementary, and that combining selection strategies may be necessary to robustly identify the full LRD population from photometry alone, albeit with a potential increase in the total number of contaminants.

Nevertheless, a substantial fraction of the RUBIES LRDs remains unrecovered by all photometric selection methods. 
Our analysis has, thus far, neglected one crucial component that spectroscopy provides over photometry: robust redshifts. 
Redshift is used explicitly in the v-shape fitting of \citet{Kocevski2024} and implicitly enters the color selection of \citet{Kokorev2024}.
These selections may therefore be influenced by uncertainties or systematics in photometric redshift estimation.

\begin{figure}
    \centering
    \includegraphics[width=\columnwidth]{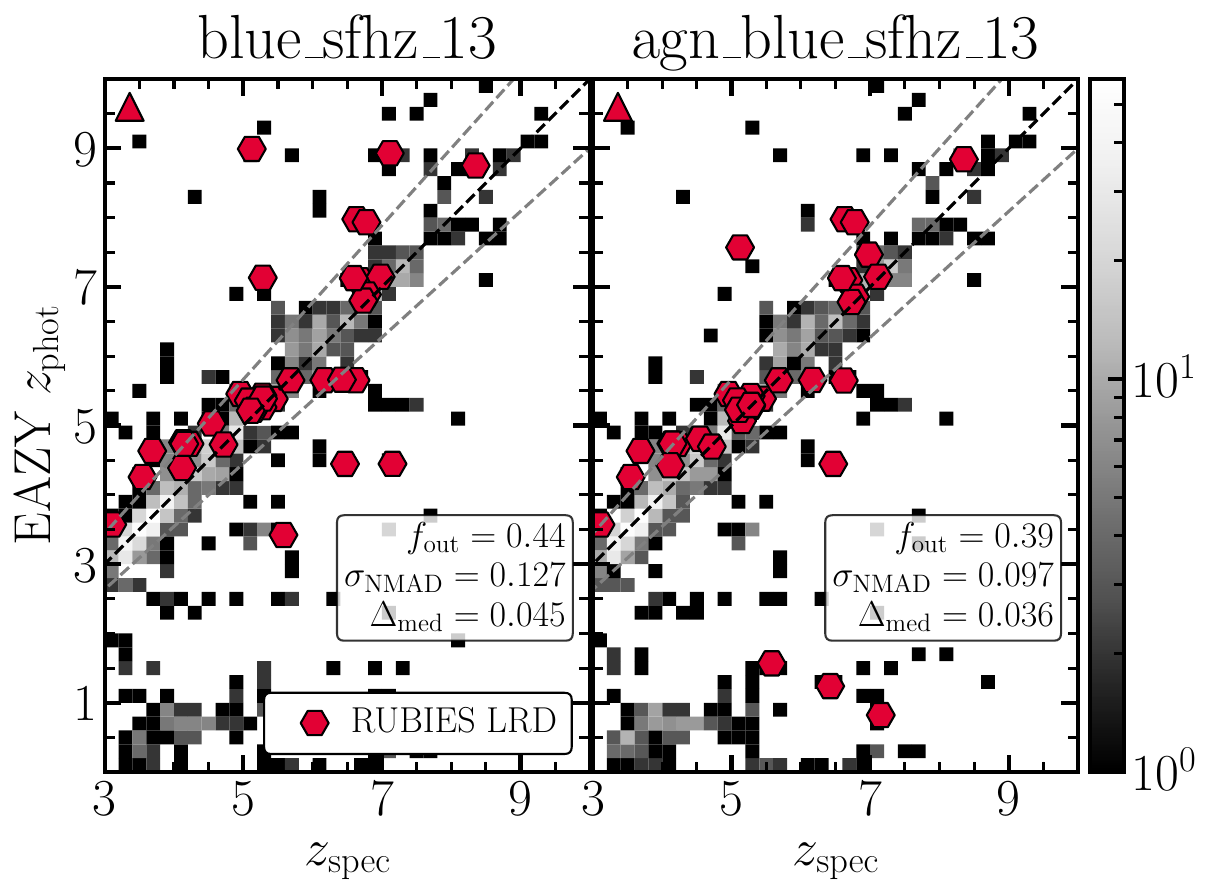}
    \caption{Best-fit photometric versus spectroscopic redshift for RUBIES sources with robust $\zspec > 3.1$ (grey histogram) and the LRD sample (red hexagons; triangle indicates out of plot). Photometric redshifts were measured using \texttt{eazy} \citep{Brammer2008} with two different template sets: the \texttt{blue\_sfhz\_13} set optimized for high-redshift galaxies (left) and the \texttt{agn\_blue\_sfhz\_13} set which additionally includes an AGN template constructed from the LRD of \citet{Killi2023} and a torus model (right). Dashed lines show $|\Delta z|/(1+\zspec) = 0.1$. Although including this AGN template reduces the number of outliers and photometric redshift scatter for the LRD sample, a small bias toward higher photometric redshifts remains. }
    \label{fig:photoz}
\end{figure}

We evaluate the photometric redshift success for the robust $\zspec>3.1$ and spectroscopic LRD samples in the left panel of Figure~\ref{fig:photoz}. 
The reddest sources in RUBIES were selected without any initial photometric redshift constraint and therefore provide an ideal comparison sample.
The photometric redshifts were obtained from template fitting to the cross-matched PSF-matched photometry from \citet[][]{Weibel2024} with \texttt{eazy} \citep{Brammer2008}, using the  \texttt{blue\_sfhz\_13} template set.
Statistics are performed on the photometric redshift deviation: $\Delta = |\Delta z|/(1+\zspec)$.
We find that the photometric redshifts of the LRDs are slightly higher than the spectroscopic redshifts (median deviation of $\Delta_\textrm{med} =0.045$) when compared with little-to-no deviation in the RUBIES $\zspec>3.1$ sample ($\Delta_\textrm{med} = -0.002$). Most importantly, we find a high outlier fraction $f_{\rm out}=0.44$ (defined as $\Delta>0.1$) and scatter ($\sigma_{\rm NMAD}=0.127$), exceeding that of the full RUBIES spectroscopic sample by a factor two and three respectively ($f_{\rm out}=0.19$  and scatter ($\sigma_{\rm NMAD}=0.034$).

Indeed, we find that the photometric redshift outliers are preferentially excluded by the photometric selections.
Of the fourteen photometric redshift outliers in the magnitude-limited LRD sample, \citet{Kocevski2024} recovers six (41\%) while \citet{Kokorev2024} recovers only one (14\%). 
Notably, all spectroscopic LRDs missed by both photometric selections are redshift outliers. 
Among these are sources with very strong Balmer breaks, such as RUBIES-UDS-154183 \citep[\emph{The Cliff}][]{deGraaff2025} and RUBIES-UDS-144195, the latter having such a strong Balmer break that it is instead fit as a Lyman break at $\zphot\sim14$ (red triangle in Figure~\ref{fig:photoz}).

Improving the completeness of future photometric selections will therefore require better treatment of the unusual colors of LRDs, which are currently not captured by commonly used template sets for photometric redshift fitting or not fully accounted for in the definition of color criteria.
Recently, a dedicated LRD template was introduced in \texttt{eazy}, which was constructed from the LRD spectrum of \citet{Killi2023} combined with a hot dusty torus model that rises sharply in the near- to mid-infrared. 
We also test the inclusion of this template (\texttt{agn\_blue\_sfhz\_13}) in the right panel Figure~\ref{fig:photoz}, and find that it sightly improves outlier fraction ($f_{\rm out}=0.39$), scatter ($\sigma_{\rm NMAD}=0.097$), and median deviation ($\Delta_\textrm{med} = 0.036$). \footnote{ 
We caution that the opposite is true for the general population of RUBIES sources, as $\Delta=-0.003$, $\sigma_{\rm NMAD}=0.037$ and $f_{\rm out}=0.18$ are all slightly worse than for the fits without an LRD template.}
However, we note that a typical torus model may fail to properly describe LRDs as many have shown to have flat mid-IR colors not well described by hot dust
\citep{Williams2024, Setton2025, wangRUBIESJWSTNIRSpec2025, deGraaff2025,chenDustBudgetCrisis2025}.

The success of photometric LRD selection hinges on accurately capturing their unique spectral properties.
Although current methods achieve high accuracy in identifying LRDs, their incompleteness underscores the limitations of relying solely on broad-band photometry. 
The high photometric redshift outlier fraction of spectroscopic LRDs, particularly among LRDs missed by both v-shaped and multi-color selection methods, suggests that standard high-redshift galaxy or AGN templates fail to fully describe LRD SEDs.
In the future, developing new empirical LRD templates and combining complementary selection criteria may maximize photometric LRD completeness.
Large spectroscopic campaigns like RUBIES will be essential to validate such photometric selections across the full color-morphology parameter space.

\section{Summary} \label{sec:summary}

We have used the large spectroscopic dataset of JWST program RUBIES to quantify the characteristics commonly associated with LRDs -- broad Balmer emission, v-shaped continua, point-like rest-optical morphologies -- among the high-redshift galaxy population. 
Because the RUBIES targets span a wide range in color and redshift without any morphological pre-selection, they are uniquely positioned to determine the prevalence of each LRD characteristic across the galaxy population.

We present one of the largest spectroscopic samples of broad Balmer line sources at high redshift to date, including 80 robust detections at $z>3.1$, 28 (35\%) of which lie at $z > 6$. 
Our detections are enabled by the \texttt{unite} package, which combines multiple NIRSpec spectra, i.e.\ the PRISM and G395M spectra, simultaneously to model emission lines and continua. 
This approach leverages the signal-to-noise of the PRISM and the resolution of G395M to identify kinematic components of emission lines, overcoming limitations of single-disperser analyses.

Moreover, we use the RUBIES spectra to measure v-shaped continua, and measure detailed morphologies from NIRCam imaging to determine which sources are point-source dominated in the rest-optical. 
Remarkably, the definition of an LRD emerges naturally from the measured spectroscopic and photometric features: all point sources with (spectroscopic) v-shaped continua exhibit broad Balmer lines where the data quality permits detection. 
This strongly suggests that these three features are not coincidental but likely stem from a common physical origin. 
Possibly, this can be explained in the context of the recently-proposed phenomenon of a massive accreting black hole embedded in an envelope of dense gas \citep{Inayoshi2025,Ji2025,Naidu2025,deGraaff2025,kidoBlackHoleEnvelopes2025,taylorCAPERSLRDz9GasEnshrouded2025}.

In detail, our primary findings can be summarized as follows:
\begin{enumerate}
    \item The sample of broad Balmer line sources span a wide range in rest-optical broad-band photometric colors, but are typically redder than the full RUBIES sample at $z>3.1$ due to a combination of red rest-optical continua and high equivalent width emission lines.
    \item The majority of sources with (spectroscopic) v-shaped continua also show a dominant point source component at rest-optical wavelengths. 
    Remarkably, we robustly detect broad Balmer emission lines for upwards of 80\% of these sources, with the remainder being inconclusive primarily due to data shortcomings. 
    Similarly, the majority of sources with v-shaped continua show broad Balmer emission lines. Of this population, 80\% also show a prominent point-source component in rest-optical imaging.
    \item Strikingly, a population of sources with all three features therefore emerges naturally from the data, and we therefore define this sample as spectroscopic LRDs, constituting the largest such sample (36) to date. 
    Compared to the full sample of broad Balmer line sources, spectroscopic LRDs are the most luminous \Ha emitters at any $\rm M_{UV}$.
    \item Photometric LRD selections are highly accurate in their recovery of broad Balmer line sources (65-80\%), as well as spectroscopic LRDs (up to $\sim$95\% for $\rm F444W<26.5$). 
    However, only $50-60\%$ of the RUBIES LRDs were previously identified in these photometric searches. 
    In particular, the selections are increasingly incomplete for bluer broad-band rest-optical colors and fainter UV magnitudes. 
    \item Combining multiple photometric LRD selections improves completeness but still leaves a large fraction of LRDs unrecovered. Sources that were missed typically have highly uncertain photometric redshifts and/or very strong Balmer breaks. This highlights the need for more nuanced photometric selection criteria as well as large spectroscopic surveys such as RUBIES.
\end{enumerate}

With a clear definition of the spectro-photometric properties that select LRDs, we are now able to build a large, robust sample of LRDs within RUBIES and the broader JWST spectroscopic archive. We can then begin to examine their detailed properties, such as the prevalence of Balmer absorption features and Balmer breaks, across the entire LRD population. 
In turn, this will provide critical clues about the evolutionary stages or environmental conditions of these systems, enabling a clearer understanding of their role in galaxy evolution and black hole growth.

\begin{acknowledgements}
REH acknowledges support by the German Aerospace Center (DLR) and the Federal Ministry for Economic Affairs and Energy (BMWi) through program 50OR2403 ‘RUBIES’.
\\
TBM was supported by a CIERA Postdoctoral Fellowship. This work used computing resources provided by Northwestern University and the Center for Interdisciplinary Exploration and Research in Astrophysics (CIERA). This research was supported in part through the computational resources and staff contributions provided for the Quest high performance computing facility at Northwestern University which is jointly supported by the Office of the Provost, the Office for Research, and Northwestern University Information Technology. 
\\
Support for this work was provided by The Brinson Foundation through a Brinson Prize Fellowship grant.
\\
The Cosmic Dawn Center is funded by the Danish National Research Foundation (DNRF) under grant \#140. This work has received funding from the Swiss State Secretariat for Education, Research and Innovation (SERI) under contract number MB22.00072, as well as from the Swiss National Science Foundation (SNSF) through project grant 200020\_207349. 
\\ 
Support for this work for RPN was provided by NASA through the NASA Hubble Fellowship grant HST-HF2-51515.001-A awarded by the Space Telescope Science Institute, which is operated by the Association of Universities for Research in Astronomy, Incorporated, under NASA contract NAS5-26555. 
\\ 
The work of CCW is supported by NOIRLab, which is managed by the Association of Universities for Research in Astronomy (AURA) under a cooperative agreement with the National Science Foundation.  
\\
This work is based in part on observations made with the NASA/ESA/CSA James Webb Space Telescope. The data were obtained from the Mikulski Archive for Space Telescopes at the Space Telescope Science Institute, which is operated by the Association of Universities for Research in Astronomy, Inc., under NASA contract NAS 5-03127 for JWST. These observations are associated with programs numbers 1345, 1837, 2234, 2279, 2514, 2750, 3990 and 4233. Support for program no.\ 4233 was provided by NASA through a grant from the Space Telescope Science Institute, which is operated by the Association of Universities for Research in Astronomy, Inc., under NASA contract NAS 5-03127. 
\\
The authors acknowledge the CEERS, PRIMER, PANORAMIC, and BEACONS teams for developing their observing program with a zero-exclusive-access period.
\\
We acknowledge the use of the following software packages which were instrumental in the development of this work: 
\texttt{Astropy} \citep{astropycollaborationAstropyCommunityPython2013,astropycollaborationAstropyProjectBuilding2018,astropycollaborationAstropyProjectSustaining2022},
\texttt{grizli} \citep{grizli},
\texttt{jax} \citep{jax2018github},
\texttt{jwst} \citep{bushouseJWSTCalibrationPipeline2022},
\LaTeX\ \citep{lamportLaTeXDocumentPreparation1994}, 
\texttt{Matplotlib} \citep{hunterMatplotlib2DGraphics2007},
\texttt{msaexp} \citep{msaexp},
\texttt{msafit} \citep{deGraaff2024a},
\texttt{NumPy} \citep{oliphantGuideNumPy2006,vanderwaltNumPyArrayStructure2011, harrisArrayProgrammingNumPy2020},
\texttt{NumPyro} \citep{phan2019}, 
\texttt{photutils} \citep{larry_bradley_2024_13989456},
\texttt{pysersic} \citep{pashaPysersicPythonPackage2023},
\texttt{photutils}
\citep{bradley_astropyphotutils_2024},
\texttt{sedpy} \citep{johnsonBdjSedpySedpy2021},
\texttt{Source-Extractor} \citep{bertinSExtractorSoftwareSource1996},
and \texttt{unite} \citep{raphael_erik_hviding_2025_15585035}.
\\
This work makes use of color palettes created by Martin Krzywinski designed for colorblindness. The color palettes and more information can be found at \url{http://mkweb.bcgsc.ca/colorblind/}.

\end{acknowledgements}

\clearpage
\bibliographystyle{aa}
\bibliography{lrds}

\appendix

\section{LRDs from RUBIES}\label{app:lrd}

We present all spectroscopically identified LRDs from RUBIES, along with all v-shaped and unresolved sources where we are unable to identify broad lines due to data quality limitations. 
We identify a total of 36 spectroscopic LRDs and seven additional sources with v-shaped continua and dominant rest-optical point sources. 
In Figure~\ref{fig:lrds} we present the color images (constructed from the F444W, F277W and F150W NIRCam filters), PRISM spectra, and G395M zoom-ins of the \Ha broad emission for all 43 objects. 
Key properties of spectroscopic LRDs and unresolved v-shaped galaxies are provided in Tables~\ref{tab:lrd} and \ref{tab:vshape-ps} respectively. We caution that the reported broad Balmer FWHM values are derived from Gaussian fitting with $\rm FWHM<2500\kms$ that may not best represent the spectral profile.
In Table~\ref{tab:vshape-ps} we also include the reason why each unresolved v-shaped galaxy has an indeterminate broad line, i.e. due to a DQ issue or missing forbidden or Balmer line.

\begin{table*}
\renewcommand{\arraystretch}{1.1}
\caption{RUBIES Spectroscopic LRDs}
\begin{center}
\begin{tabular}{crrrcrp{1.4em}p{0em}lp{1.75em}p{0em}lp{1.15em}p{0em}lc}
\hline
\hline
    Field & 
    \multicolumn{1}{c}{ID} & 
    \multicolumn{1}{c}{R.A.} & 
    \multicolumn{1}{c}{Dec} & 
    $\zspec$\textsuperscript{a} & 
    $\rm \Delta WAIC$ &
    \multicolumn{3}{c}{Broad FWHM\textsuperscript{b}} & 
    \multicolumn{3}{c}{$\beta_\textrm{UV}$} & 
    \multicolumn{3}{c}{$\beta_\textrm{opt.}$} & 
    F444W\textsuperscript{c} 
\\
    & 
    \multicolumn{1}{c}{} &
    \multicolumn{2}{c}{J2000} 
    & 
    & 
    &
    \multicolumn{3}{c}{$\kms$} & 
    \multicolumn{3}{c}{} &
    \multicolumn{3}{c}{} &
    \multicolumn{1}{c}{mag} 
\\
\hline
UDS & 40579 & 34.244200 & $-$5.245871 & 3.11 & 84 & 2466 & $\pm$ & 31 & $-$0.37 & $\pm$ & 0.02 & 1.97 & $\pm$ & 0.10 & 21.95 \\
UDS & 144195 & 34.325156 & $-$5.143685 & 3.36 & 143 & 2053 & $\pm$ & 130 & $-$1.52 & $\pm$ & 0.08 & 0.64 & $\pm$ & 0.02 & 24.52 \\
UDS & 154183 & 34.410749 & $-$5.129664 & 3.55 & 367 & 2481 & $\pm$ & 18 & $-$0.69 & $\pm$ & 0.07 & 1.12 & $\pm$ & 0.01 & 23.23 \\
UDS & 23438 & 34.338271 & $-$5.280895 & 3.69 & 69 & 2058 & $\pm$ & 183 & $-$1.79 & $\pm$ & 0.04 & 0.64 & $\pm$ & 0.05 & 24.71 \\
UDS & 167741 & 34.334603 & $-$5.110269 & 4.12 & 25 & 2399 & $\pm$ & 99 & $-$1.88 & $\pm$ & 0.04 & 0.66 & $\pm$ & 0.19 & 24.69 \\
UDS & 31747 & 34.223757 & $-$5.260245 & 4.13 & 98 & 2277 & $\pm$ & 150 & $-$0.91 & $\pm$ & 0.19 & 0.85 & $\pm$ & 0.01 & 24.80 \\
UDS & 119957 & 34.268908 & $-$5.176722 & 4.15 & 194 & 1978 & $\pm$ & 89 & $-$1.66 & $\pm$ & 0.06 & 0.91 & $\pm$ & 0.03 & 25.40 \\
EGS & 28812 & 214.924149 & 52.849050 & 4.22 & 841 & 2194 & $\pm$ & 60 & $-$0.61 & $\pm$ & 0.08 & 1.04 & $\pm$ & 0.03 & 24.47 \\
EGS & 29489 & 215.022071 & 52.920786 & 4.54 & 207 & 2370 & $\pm$ & 85 & $-$1.63 & $\pm$ & 0.13 & 0.80 & $\pm$ & 0.10 & 25.84 \\
UDS & 182791 & 34.213813 & $-$5.087050 & 4.72 & 1123 & 2496 & $\pm$ & 4 & $-$1.23 & $\pm$ & 0.01 & 0.43 & $\pm$ & 0.03 & 24.86 \\
EGS & 42232 & 214.886792 & 52.855381 & 4.95 & 393 & 2449 & $\pm$ & 40 & $-$1.00 & $\pm$ & 0.07 & 0.42 & $\pm$ & 0.03 & 22.74 \\
EGS & 61496 & 214.972441 & 52.962192 & 5.08 & 40 & 1509 & $\pm$ & 173 & $-$1.06 & $\pm$ & 0.07 & 0.89 & $\pm$ & 0.04 & 25.82 \\
EGS & 952625 & 214.975529 & 52.925268 & 5.11 & 28 & 1367 & $\pm$ & 337 & $-$1.71 & $\pm$ & 0.37 & 2.31 & $\pm$ & 0.31 & 26.40 \\
UDS & 830237 & 34.374852 & $-$5.275529 & 5.12 & 57 & 1238 & $\pm$ & 214 & $-$2.50 & $\pm$ & 0.13 & 0.96 & $\pm$ & 0.26 & 26.53 \\
UDS & 149298 & 34.424311 & $-$5.136491 & 5.15 & 65 & 2120 & $\pm$ & 248 & $-$1.07 & $\pm$ & 0.03$^\dagger$ & 3.41 & $\pm$ & 0.60 & 24.23 \\
UDS & 53692 & 34.455376 & $-$5.231814 & 5.28 & 43 & 1725 & $\pm$ & 301 & $-$2.25 & $\pm$ & 0.43 & 1.46 & $\pm$ & 0.09 & 26.26 \\
EGS & 42046 & 214.795368 & 52.788847 & 5.28 & 4558 & 2499 & $\pm$ & 1 & $-$0.23 & $\pm$ & 0.01 & 0.65 & $\pm$ & 0.03 & 23.16 \\
UDS & 970351 & 34.261900 & $-$5.105205 & 5.28 & 51 & 2396 & $\pm$ & 91 & $-$0.84 & $\pm$ & 0.07$^\dagger$ & 0.76 & $\pm$ & 0.05 & 25.64 \\
EGS & 926125 & 215.137081 & 52.988554 & 5.28 & 1426 & 1949 & $\pm$ & 57 & $-$1.09 & $\pm$ & 0.04 & 1.93 & $\pm$ & 0.05 & 25.26 \\
UDS & 29813 & 34.453355 & $-$5.270717 & 5.44 & 477 & 2245 & $\pm$ & 101 & $-$2.27 & $\pm$ & 0.13 & 0.98 & $\pm$ & 0.08 & 25.80 \\
UDS & 172350 & 34.368951 & $-$5.103941 & 5.58 & 592 & 1922 & $\pm$ & 46 & $-$1.36 & $\pm$ & 0.04 & 0.37 & $\pm$ & 0.02 & 24.57 \\
EGS & 37124 & 214.990977 & 52.916524 & 5.68 & 145 & 2453 & $\pm$ & 44 & $-$2.03 & $\pm$ & 0.26$^\dagger$ & 0.80 & $\pm$ & 0.01 & 25.13 \\
UDS & 50716 & 34.313154 & $-$5.226781 & 6.17 & 61 & 2250 & $\pm$ & 159 & $-$1.61 & $\pm$ & 0.12 & 2.25 & $\pm$ & 0.13 & 25.80 \\
UDS & 57040 & 34.470042 & $-$5.238069 & 6.42 & 1567 & 2290 & $\pm$ & 34 & $-$0.84 & $\pm$ & 0.20 & 0.72 & $\pm$ & 0.01 & 24.07 \\
UDS & 36171 & 34.345003 & $-$5.260115 & 6.47 & 16 & 2177 & $\pm$ & 261 & $-$1.78 & $\pm$ & 0.35 & 0.46 & $\pm$ & 0.08 & 26.09 \\
UDS & 967770 & 34.228526 & $-$5.101526 & 6.59 & 13 & 2330 & $\pm$ & 145 & $-$1.82 & $\pm$ & 0.13 & 0.36 & $\pm$ & 0.05 & 26.04 \\
EGS & 53254 & 214.797537 & 52.818746 & 6.62 & 64 & 2067 & $\pm$ & 303 & $-$1.33 & $\pm$ & 0.01$^\dagger$ & 0.71 & $\pm$ & 0.23 & 25.73 \\
UDS & 33938 & 34.266442 & $-$5.256391 & 6.63 & 32 & 1143 & $\pm$ & 297 & $-$1.87 & $\pm$ & 0.09 & 0.98 & $\pm$ & 0.26 & 26.71 \\
EGS & 49140 & 214.892248 & 52.877410 & 6.68 & 5279 & 2499 & $\pm$ & 1 & $-$0.69 & $\pm$ & 0.01 & 1.10 & $\pm$ & 0.02 & 23.92 \\
EGS & 948917 & 214.892479 & 52.856890 & 6.73 & 292 & 2217 & $\pm$ & 134 & $-$1.21 & $\pm$ & 0.01 & 0.86 & $\pm$ & 0.02 & 25.50 \\
UDS & 807469 & 34.376139 & $-$5.310366 & 6.78 & 788 & 1742 & $\pm$ & 74 & $-$1.13 & $\pm$ & 0.39$^\dagger$ & 2.48 & $\pm$ & 0.35 & 25.03 \\
EGS & 927271 & 215.078259 & 52.948497 & 6.78 & 19 & 1649 & $\pm$ & 357 & $-$1.64 & $\pm$ & 0.02 & 0.91 & $\pm$ & 0.09 & 26.19 \\
EGS & 55604 & 214.983026 & 52.956001 & 6.98 & 6966 & 2499 & $\pm$ & 1 & $-$0.96 & $\pm$ & 0.01 & 1.50 & $\pm$ & 0.02 & 23.82 \\
UDS & 848745 & 34.310575 & $-$5.249898 & 7.11 & 86 & 1735 & $\pm$ & 292 & $-$1.12 & $\pm$ & 0.08 & 1.72 & $\pm$ & 0.13 & 26.11 \\
EGS & 42803 & 214.929524 & 52.887919 & 7.15 & 237 & 2393 & $\pm$ & 78 & $-$1.19 & $\pm$ & 0.11 & 1.67 & $\pm$ & 0.04 & 26.04 \\
EGS & 966323 & 214.876149 & 52.880831 & 8.35 & 55 & 2199 & $\pm$ & 223 & $-$1.13 & $\pm$ & 0.01 & 0.57 & $\pm$ & 0.07 & 25.68 \\
\hline
\hline
\end{tabular}
\label{tab:lrd}
\end{center}
\tablefoot{The machine-readable version of this table is available on \href{https://doi.org/10.5281/zenodo.15528783}{Zenodo} \citep{hviding_2025_15528783}.\\
\textsuperscript{a}Based on the best DJA value.\\
\textsuperscript{b}We caution that these values are derived from Gaussian fitting with $\rm FWHM<2500\kms$ that may not best represent the spectral profile.\\
\textsuperscript{c}Obtained from the \citet{Weibel2024} catalogs.\\
$^\dagger\beta_\textrm{UV}$ derived from photometry.
}
\end{table*}

\begin{table*}
\renewcommand{\arraystretch}{1.1}
\caption{RUBIES V-Shaped Rest-Optical Point Sources (with Indeterminate Broad-Balmer Emission)}
\begin{center}
\begin{tabular}{crrrccp{1.75em}p{0em}lp{1.15em}p{0em}lc}
\hline
\hline
    Field & 
    \multicolumn{1}{c}{ID} & 
    \multicolumn{1}{c}{R.A.} & 
    \multicolumn{1}{c}{Dec} & 
    $\zspec$\textsuperscript{a} & 
    \multicolumn{1}{c}{Broad H\,{\sc i}\xspace Note} & 
    \multicolumn{3}{c}{$\beta_\textrm{UV}$} & 
    \multicolumn{3}{c}{$\beta_\textrm{opt.}$} & 
    F444W\textsuperscript{b}
\\
    & 
    \multicolumn{1}{c}{} &
    \multicolumn{2}{c}{J2000} 
    & 
    & 
    \multicolumn{1}{c}{} & 
    \multicolumn{3}{c}{} &
    \multicolumn{3}{c}{} &
    \multicolumn{1}{c}{mag} 
\\
\hline
UDS & 176292 & 34.228100 & $-$5.098144 & 3.14 & Edge of Range & $-$2.58 & $\pm$ & 0.86 & 1.48 & $\pm$ & 0.09 & 25.27 \\
UDS & 16053 & 34.367104 & $-$5.293524 & 3.95 & No \Oiii & $-$1.37 & $\pm$ & 0.40 & 0.60 & $\pm$ & 0.05 & 25.50 \\
UDS & 147411 & 34.360718 & $-$5.139081 & 3.97 & No [O\,{\sc iii}] & $-$1.24 & $\pm$ & 0.08 & 0.17 & $\pm$ & 0.05 & 26.29 \\
UDS & 40800 & 34.324122 & $-$5.252789 & 4.98 & \Hb \& \Oiii DQ & $-$2.51 & $\pm$ & 0.98$^\dagger$ & 2.52 & $\pm$ & 0.25 & 26.67 \\
EGS & 69475 & 214.949187 & 52.964143 & 5.62 & No \Ha & $-$1.73 & $\pm$ & 0.03 & 0.37 & $\pm$ & 0.09 & 25.76 \\
UDS & 928474 & 34.243700 & $-$5.271438 & 6.93 & Contamination & $-$2.28 & $\pm$ & 0.22 & 1.55 & $\pm$ & 0.22 & 26.20 \\
EGS & 932920 & 214.927819 & 52.850001 & 7.48 & No \Ha & $-$1.97 & $\pm$ & 0.01 & 0.69 & $\pm$ & 0.14 & 26.69 \\
\hline
\hline
\end{tabular}
\label{tab:vshape-ps}
\end{center}
\tablefoot{The machine-readable version of this table is available on \href{https://doi.org/10.5281/zenodo.15528783}{Zenodo} \citep{hviding_2025_15528783}.\\
\textsuperscript{a}Based on the best DJA value.\\
\textsuperscript{b}Obtained from the \citet{Weibel2024} catalogs.\\
$^\dagger\beta_\textrm{UV}$ derived from photometry.
}
\end{table*}

\begin{figure*}[!ht]
    \centering
    \includegraphics[width=\textwidth]{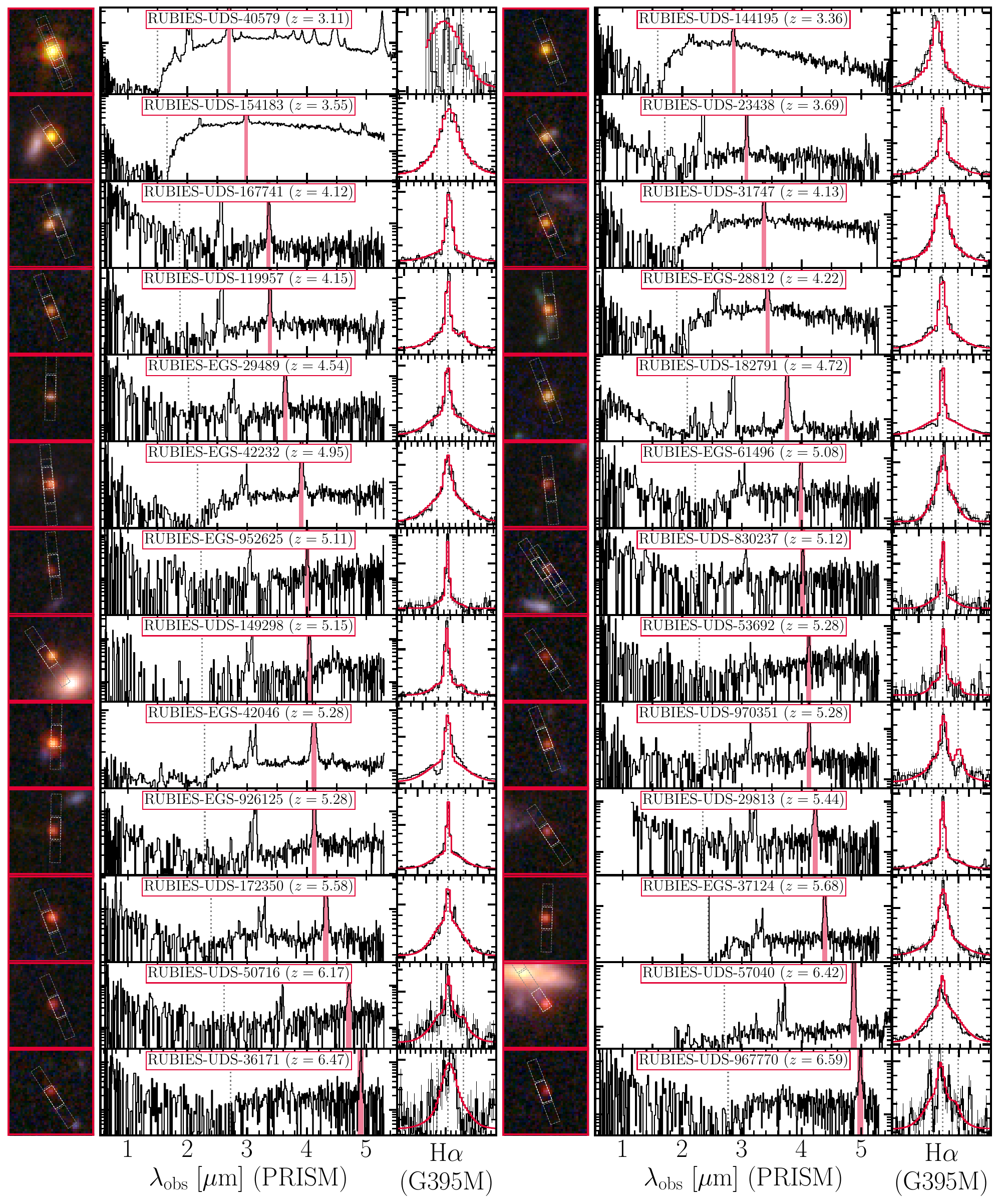}
    \caption{All RUBIES LRDs (red) found in this work followed by rest-optical point sources with v-shaped continua (orange) all organized by redshift. In each column in each row we show the following: the 1"$\times$1" F444W/F277W/F150W RGB cutouts (left), the log-scaled PRISM spectra with the location of the Balmer limit marked with a dashed line (middle), and the linear-scaled G395M $\pm3000\kms$ zoom-in of the \Ha line where available, otherwise \Hb, with the broad \texttt{unite} fit superimposed (right).}
    \label{fig:lrds}
\end{figure*}
\begin{figure*}[!ht]
    \ContinuedFloat
    \centering    \includegraphics[width=\textwidth]{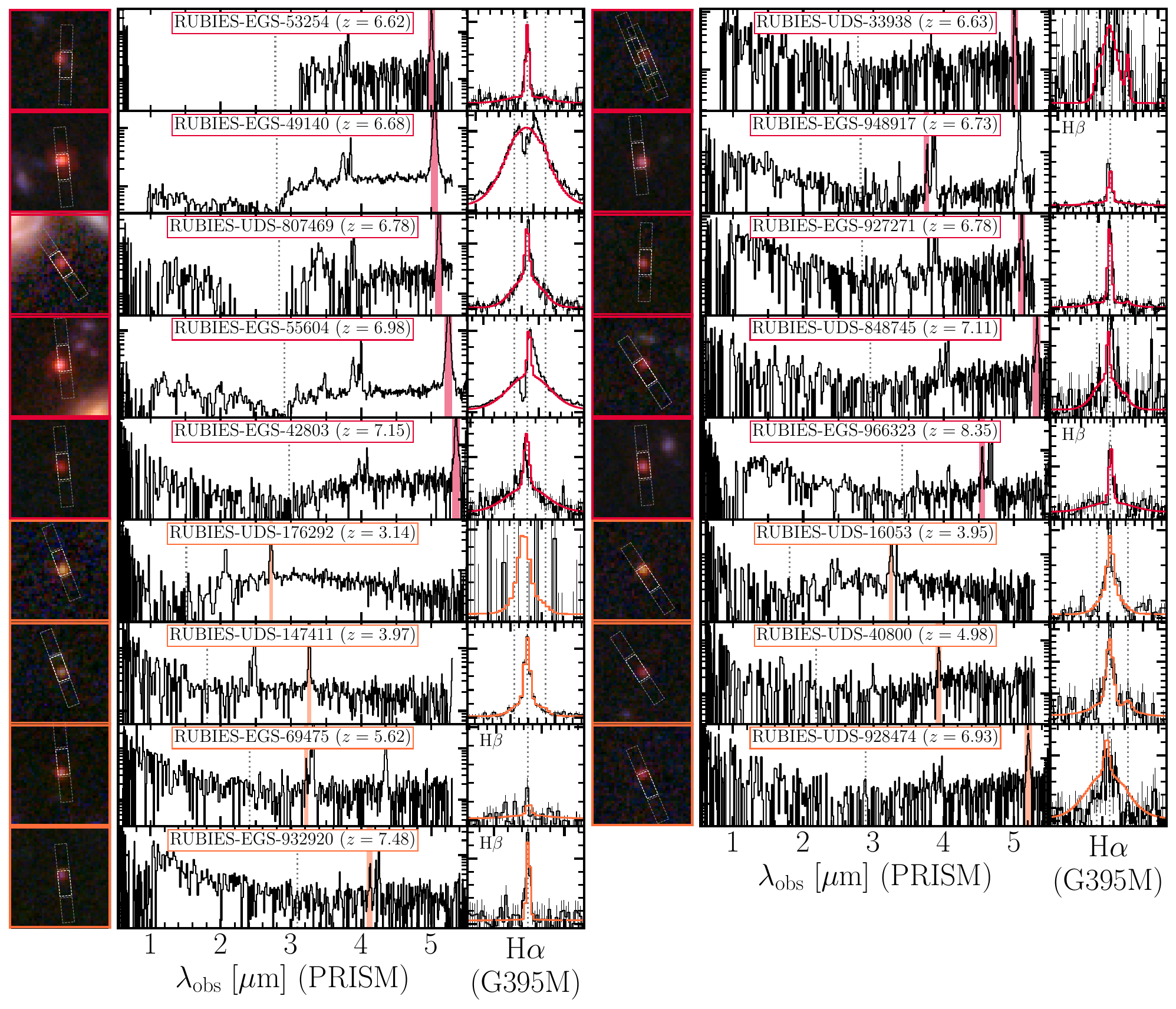}
    \caption{All RUBIES LRDs (red) found in this work followed by rest-optical point sources with v-shaped continua (orange) all organized by redshift. In each column in each row we show the following: the 1"$\times$1" F444W/F277W/F150W RGB cutouts (left), the log-scaled PRISM spectra with the location of the Balmer limit marked with a dashed line (middle), and the linear-scaled G395M $\pm3000\kms$ zoom-in of the \Ha line where available, otherwise \Hb, with the broad \texttt{unite} fit superimposed (right) (continued).}
\end{figure*}

\section{Broad Line Sample} \label{app:broad}

\begin{figure*}[!ht]
    \centering
    \includegraphics[width=\textwidth]{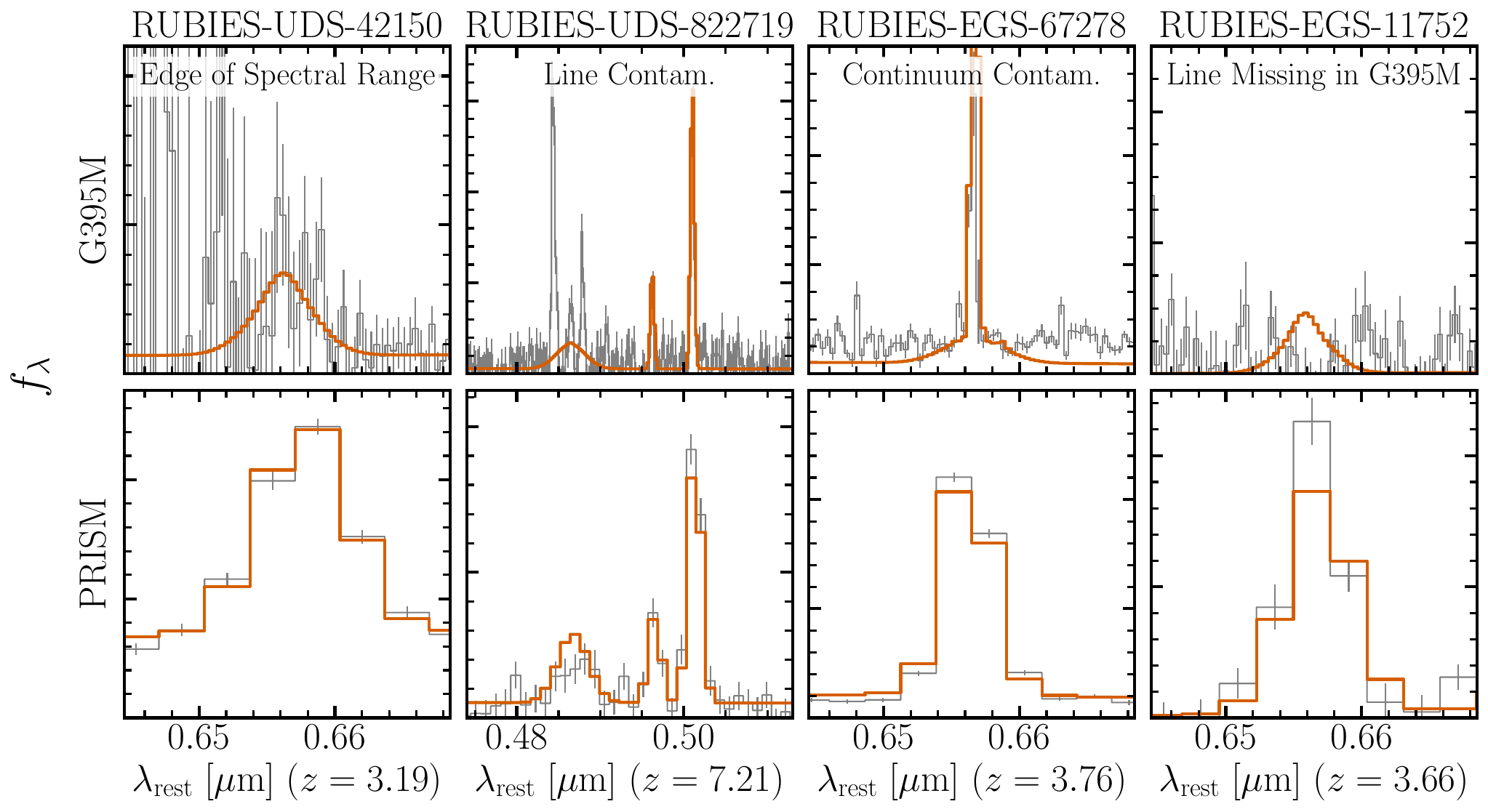}
    \caption{Examples of failed fits due to artifacts at the edge of the spectral range, emission line contamination from trace overlaps, continuum contamination from trace overlaps, and inconsistencies between G395M and PRISM extractions for RUBIES-UDS-42150, RUBIES-UDS-822719, RUBIES-EGS-67278, and RUBIES-EGS-11752 respectively.}
    \label{fig:badfit}
\end{figure*}

We present the results of our broad-line fitting, including both successful detections and cases where our fitting procedure failed due to data quality issues. 
In Figure~\ref{fig:badfit} we show how trace overlaps in G395M or other data quality artifacts can produce a bad fit with a broad line:
\begin{enumerate}
    \item RUBIES-UDS-42150: Here the \Ha emission falls at the edge of the G395M spectral range, leading to data quality issues.
    \item RUBIES-UDS-822719: Emission line contamination from an adjacent trace can artificially induce flux in the G395M that can be better fit with a broad component.
    \item RUBIES-EGS-67278: Trace overlap may also manifest as additional continuum flux in the G395M spectrum, leading to a bad fit with a broad line.
    \item RUBIES-EGS-11752: In rare cases, a line might be missing entirely from the G395M data
\end{enumerate}

\begin{figure*}[!ht]
    \centering
    \includegraphics[width=\textwidth]{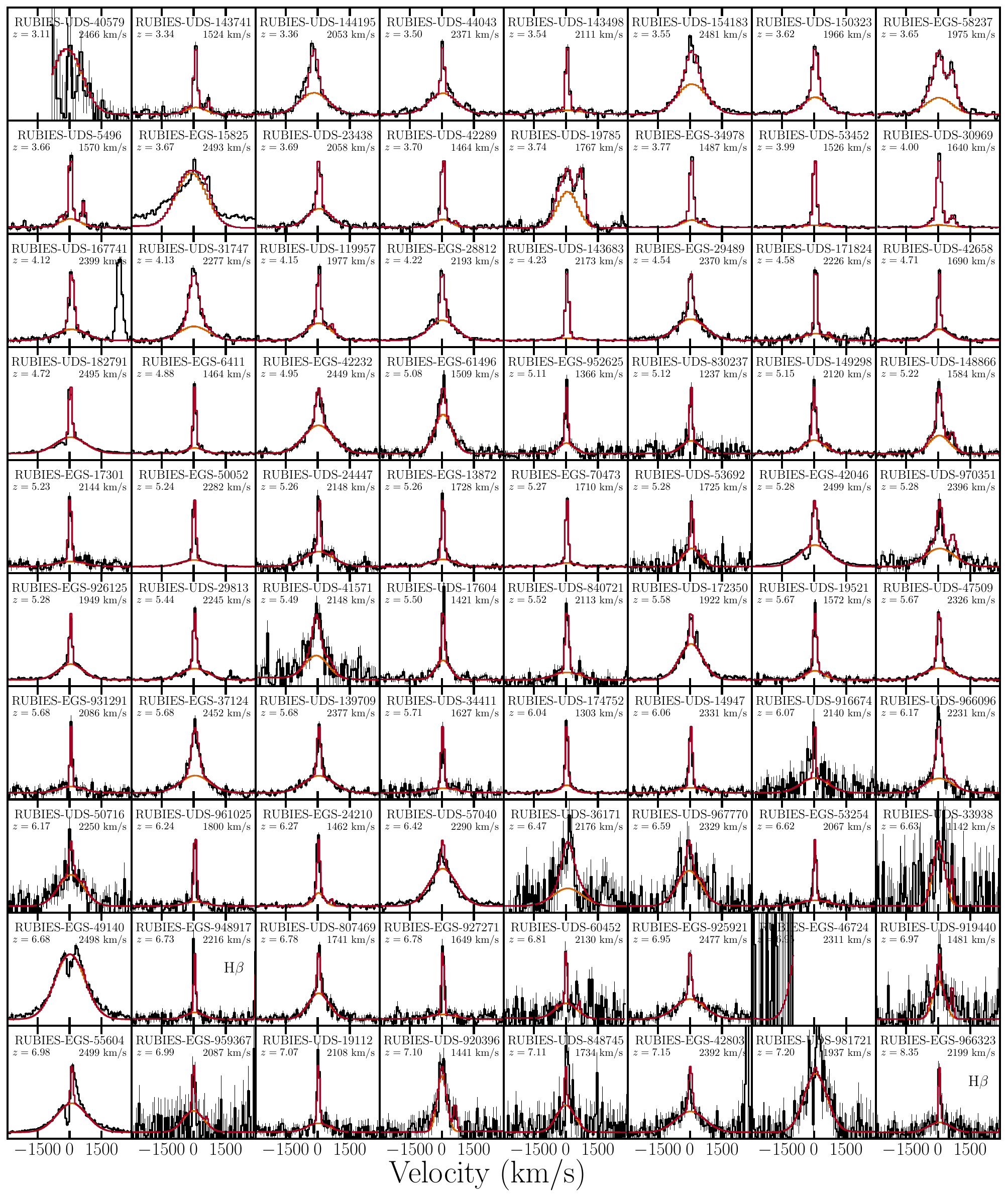}
    \caption{
    All robust Broad-Balmer line objects found in the RUBIES data, organized by redshift. 
    We plot the linear-scaled G395M $\pm3000\kms$ zoom-in of the \Ha line where available, otherwise \Hb, with the broad \texttt{unite} fit superimposed in red and the broad Balmer component in orange. We also include text containing the FWHM of the broad component as well.}
    \label{fig:bl}
\end{figure*}

Our analysis identified 80 robust broad Balmer line sources and 18 potential broad-line systems
The complete catalog of broad-line sources is provided in Table~\ref{tab:bl}, while Figure~\ref{fig:bl} shows spectral zoom-ins for all robust and potential broad-line systems.
However, we caution that the reported broad Balmer FWHM values are derived from Gaussian fitting with $\rm FWHM<2500\kms$, optimized for broad line detection, that may not best represent the spectral profile.

\begin{table*}
\renewcommand{\arraystretch}{1.1}
\caption{RUBIES $\zspec > 3.1$ Broad-Balmer Line Galaxies}
\begin{center}
\begin{tabular}{crrrcrp{1.4em}p{0em}lc}
\hline
\hline
    Field & 
    \multicolumn{1}{c}{ID} & 
    \multicolumn{1}{c}{R.A.} & 
    \multicolumn{1}{c}{Dec} & 
    $\zspec^\textrm{a}$ & 
    $\rm\Delta WAIC$ &
    \multicolumn{3}{c}{Broad FWHM\textsuperscript{b}} & 
    F444W\textsuperscript{c}
\\
    & 
    \multicolumn{1}{c}{} &
    \multicolumn{2}{c}{J2000} 
    &
    & 
    & 
    \multicolumn{3}{c}{$\kms$} & 
    \multicolumn{1}{c}{mag} 
\\
\hline
UDS & 40579 & 34.244200 & $-$5.245871 & 3.11 & 84 & 2466 & $\pm$ & 31 & 21.95 \\
UDS & 143741 & 34.330086 & $-$5.144587 & 3.34 & 13 & 1524 & $\pm$ & 409 & 23.39 \\
UDS & 144195 & 34.325156 & $-$5.143685 & 3.36 & 143 & 2053 & $\pm$ & 130 & 24.52 \\
UDS & 44043 & 34.241817 & $-$5.239436 & 3.50 & 368 & 2371 & $\pm$ & 91 & 24.47 \\
UDS & 143498 & 34.417352 & $-$5.145051 & 3.54 & 13 & 2112 & $\pm$ & 343 & 23.55 \\
UDS & 154183 & 34.410749 & $-$5.129664 & 3.55 & 367 & 2481 & $\pm$ & 18 & 23.23 \\
UDS & 150323 & 34.417822 & $-$5.134842 & 3.62 & 74 & 1966 & $\pm$ & 181 & 25.21 \\
EGS & 58237 & 214.850571 & 52.866030 & 3.65 & 32 & 1976 & $\pm$ & 142 & 22.46 \\
UDS & 5496 & 34.405872 & $-$5.312951 & 3.66 & 72 & 1570 & $\pm$ & 208 & 24.14 \\
EGS & 15825 & 215.079264 & 52.934252 & 3.67 & 303 & 2493 & $\pm$ & 7 & 21.94 \\
\multicolumn{9}{c}{$\cdots$}\\
\hline
\hline
\end{tabular}
\label{tab:bl}
\end{center}
\tablefoot{The full and machine-readable version of this table is available on \href{https://doi.org/10.5281/zenodo.15528783}{Zenodo} \citep{hviding_2025_15528783}.\\
\textsuperscript{a}Based on the best DJA value.\\
\textsuperscript{b}We caution that these values are derived from Gaussian fitting with $\rm FWHM<2500\kms$ that may not best represent the spectral profile.\\
\textsuperscript{c}Obtained from the \citet{Weibel2024} catalogs.}
\end{table*}

\end{document}